\documentclass[aps,prb,superscriptaddress,reprint,floatfix,longbibliography,amsfonts,amsmath,amssymb]{revtex4-2}
\usepackage{graphicx}
\usepackage{hyperref}
\usepackage{epstopdf}
\usepackage[note-name=, use-sort-key = false]{notes2bib}
\usepackage{color}
\usepackage{pifont}
\usepackage{array}
\usepackage{pdfpages}
\usepackage[version=4]{mhchem}
\usepackage{placeins}

\hypersetup{
    colorlinks = true,
    citecolor = blue,
    bookmarksnumbered = true,
    urlcolor = blue
}

\makeatletter
\AtBeginDocument{\let\LS@rot\@undefined}
\makeatother

\newcommand{\ch}{\ce{CsBaCr3F12}}
\newcommand{\fe}{\ce{CsBaFe3F12}}

\newcommand{\suppSecEnergyMapping}{Supplementary Note 1}
\newcommand{\suppSecMonteCarlo}{Supplementary Note 2}
\newcommand{\suppSecWavevectors}{Supplementary Note 3}
\newcommand{\suppSecStiffnessModel}{Supplementary Note 4}

\begin{document}

\title{Realization of the Ruby Lattice Antiferromagnet in Layered Transition-Metal Fluorides}

\author{Harald O. Jeschke}
\email{jeschke@okayama-u.ac.jp}
\affiliation{Research Institute for Interdisciplinary Science, Okayama University, Okayama 700-8530, Japan}

\author{Daniel Guterding}
\email{daniel.guterding@th-brandenburg.de}
\affiliation{Technische Hochschule Brandenburg, Magdeburger Straße 50, 14770 Brandenburg an der Havel, Germany}

\author{Pratyay Ghosh}
\email{pratyay.ghosh@epfl.ch}
\affiliation{Institute of Physics, Ecole Polytechnique F\'ed\'erale de Lausanne (EPFL), CH-1015 Lausanne, Switzerland}

\begin{abstract}
The antiferromagnet on the ruby lattice is expected to host a range of exotic emergent phenomena, yet its material realization has remained elusive.
Here we show that the layered transition metal fluorides {\fe} and {\ch} with Fe$^{3+}$ and Cr$^{3+}$ ions realize only slightly distorted ruby lattice geometries with spin moments $S=5/2$ and $S=3/2$, respectively. Their microscopic Hamiltonians, calculated with DFT energy mapping, are dominated by short-ranged antiferromagnetic interactions within the ruby layers. Classical Monte Carlo simulations reveal strong frustration in both compounds, with local N\'eel correlations on the hexagonal plaquettes and distinct long-range ordering tendencies governed by weaker triangular links. For {\fe}, the calculated thermodynamic behaviour is consistent with the experimentally reported magnetic ordering scale. For {\ch}, classical Monte Carlo and Luttinger-Tisza analysis reveal competing low-energy ordering wave vectors, strong finite-size sensitivity, and a tendency toward incommensurate order. Overall, our results establish these fluorides as experimentally accessible ruby-lattice antiferromagnets and provide quantitative predictions for future neutron-scattering studies.
\end{abstract}

\maketitle

\section{Introduction}
The enduring interest in frustrated quantum and classical magnets stems from their ability to realize an exceptional diversity of collective states, ranging from unconventional symmetry-broken phases~\cite{Mila2011,Diep2013} to symmetry-preserving states with emergent gauge structures~\cite{Gingras2014,Sachdev2016} and topological orders~\cite{Balents2010,Savary2016}.
In this context, the kagome and pyrochlore antiferromagnets have long served as paradigmatic frustrated spin models playing a central role in shaping modern understanding of frustrated magnetism.
The nearest-neighbour Heisenberg antiferromagnet on the kagome lattice exhibits an extensively degenerate classical manifold and strong fluctuation effects~\cite{Harris1992, Chalker1992}, while the pyrochlore counterpart realizes a classical spin-liquid regime with algebraic correlations and emergent gauge physics~\cite{Moessner1998}. 
Their relevance has been reinforced by experimental realizations,
from the kagome system herbertsmithite~\cite{Shores2005, Han2012} to magnetic pyrochlore oxides~\cite{Gardner2010}, which have established these systems as benchmark platforms for exploring frustration-driven phenomena. 

\begin{figure}[t]
\includegraphics*[width=0.9\columnwidth]{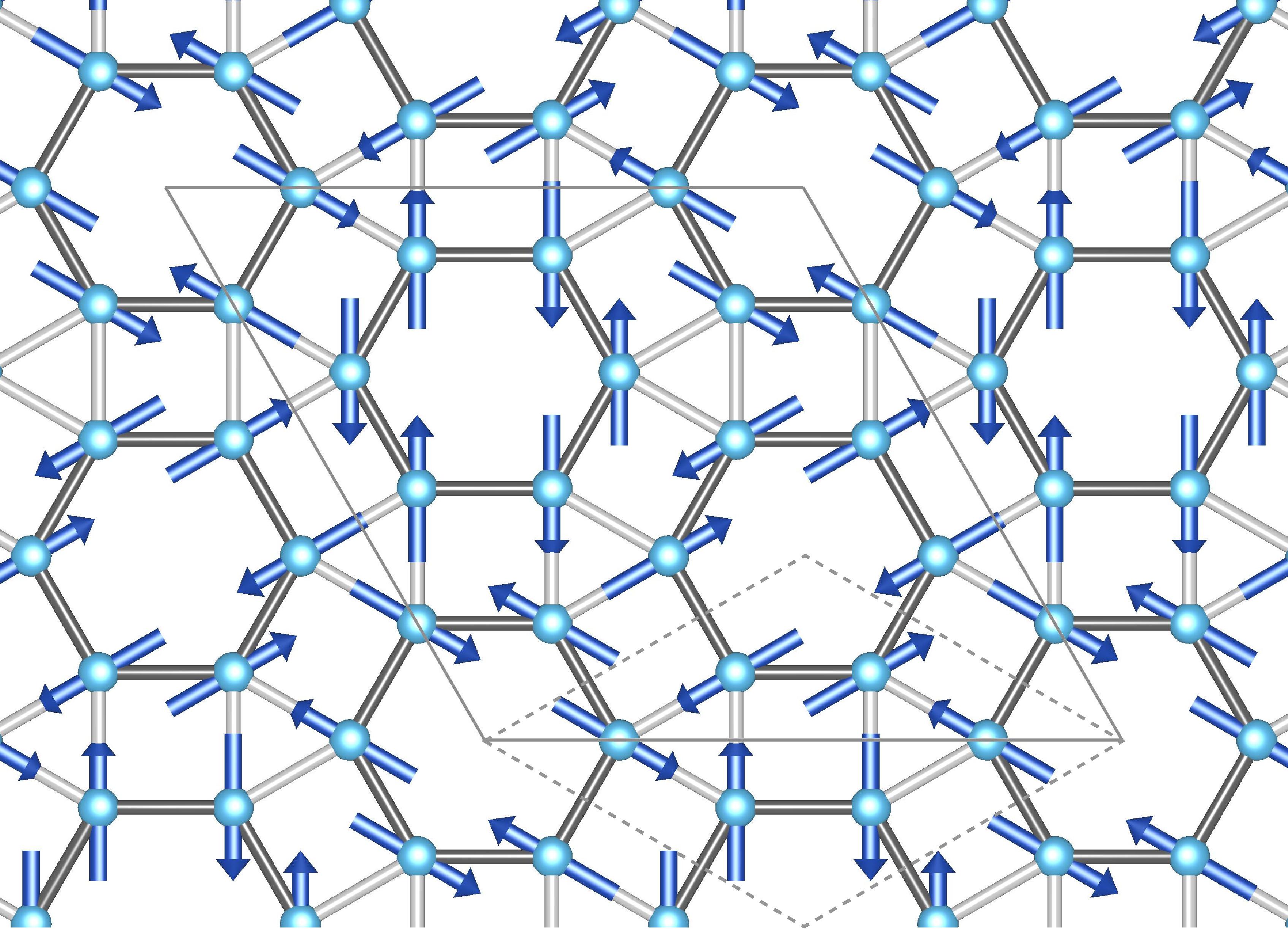}
\caption{\textbf{Ruby lattice and the ground state of the classical Heisenberg antiferromagnet.} The ideal ruby lattice geometry constructed from a periodic tiling of triangles, squares, and hexagons. The unit cell is indicated by dashed lines. The lattice contains two symmetry-inequivalent nearest-neighbour bonds, one set forming triangular motifs and another forming hexagonal loops. The classical antiferromagnetic Heisenberg model on this lattice has a ground state in which each hexagon develops local N\'eel order, with a relative $120^\circ$ spin canting between neighbouring hexagons.
}
\label{fig:ruby}
\end{figure}

Motivated by both the discovery of new materials and advances in theoretical many-body physics, recent years have witnessed a rapid expansion of interest toward a broader class of frustrated lattices in two and three dimensions.
In two dimensions, notable examples include the maple leaf lattice~\cite{Beck2024,Gembe2024}, square-kagome lattice~\cite{Niggemann2023}, star lattice~\cite{Ghosh2025b}, and trellis lattice~\cite{Chatterjee2026}. In three dimensions, prominent examples include the hyperkagome lattice~\cite{Okamoto2007,Chillal2020} and the trillium lattice~\cite{Zivkovic2021}. 
Within this expanding landscape, the ruby lattice has emerged as an especially intriguing geometry with an Archimedean tiling made of triangles, squares, and hexagons, belonging to $p6m$ symmetry (see Fig.~\ref{fig:ruby}). 
Another way to view the lattice is as a honeycomb arrangement of triangles. 
Its name was coined in 1983 in the context of solving the 2D Ising model on this lattice~\cite{Lin1983}. It has subsequently been given another nickname, the bounce lattice, in the context of percolation transitions~\cite{Suding1999}. 
Due to the presence of triangles, the antiferromagnetic spin models on the ruby lattice are intrinsically frustrated; the Ising model on this lattice is shown to possess a spin ice ground state with extensive degeneracy~\cite{Archi_Ising}. In contrast, the classical Heisenberg model on the same lattice assumes a magnetically ordered ground state in which each triangle develops a local $120^\circ$ spin configuration, while spins associated with the hexagons remain antiparallel (see Fig.~\ref{fig:ruby})~\cite{Farnell2011,Ghosh2022}. An equivalent description of this state is that the hexagons form perfect N\'eel-ordered units, with a relative $120^\circ$ canting between spins belonging to adjacent hexagons.
Apart from the classical physics, this lattice has also remained highly relevant in several contexts of purely theoretical interest, including the study of anyonic excitations~\cite{Bombin2009,Verresen2021,Verresen2022}, topological properties~\cite{Joseph2025,Yang2023,Hu2011,Semeghini2021}, and quantum magnetism~\cite{Richter2004,Farnell2011,Beck2024,Maity2024,Semeghini2021,Jahromi2016,Jahromi2018,Lukin2024,Schafer2023,ghosh2025simplexcrystalgroundstate}. In addition, the ruby lattice has recently gained attention in Rydberg atom platforms, where its nontrivial connectivity and geometric frustration provide a powerful route toward engineering strongly correlated quantum many-body states in controllable experimental settings~\cite{Ryd1,Ryd2,Ryd3,Ryd4}.

Despite this broad interest, experimental realizations of the ruby lattice remain remarkably scarce; reports of candidate ruby-lattice magnets are rare~\cite{Guo2025}, and reliable microscopic Hamiltonian characterization is unavailable in nearly all of them. In this work, we focus on two layered fluoride compounds, namely, the iron fluoride {\fe}~\cite{Renaudin1991} and the chromium fluoride {\ch}~\cite{Ferey1989}, which are arguably the only genuine material realizations of the ruby lattice to date. Compounds like \ce{YNi2Al3}~\cite{Sorgic1998} or \ce{BaEu6(Si3B6O24)(OH)2}~\cite{Heyward2015} that have been put forward in Ref.~\onlinecite{Guo2025} are in fact rather poor ruby lattice candidates. In these materials, the separation between adjacent ruby-like layers is sufficiently small that interlayer couplings are expected to be substantial. Consequently, even if they are magnetic, the resulting three-dimensional magnetism is expected to dominate the low-energy physics and obscure the characteristics of the ruby lattice.

\begin{figure*}[t]
\includegraphics*[width=0.95\textwidth]{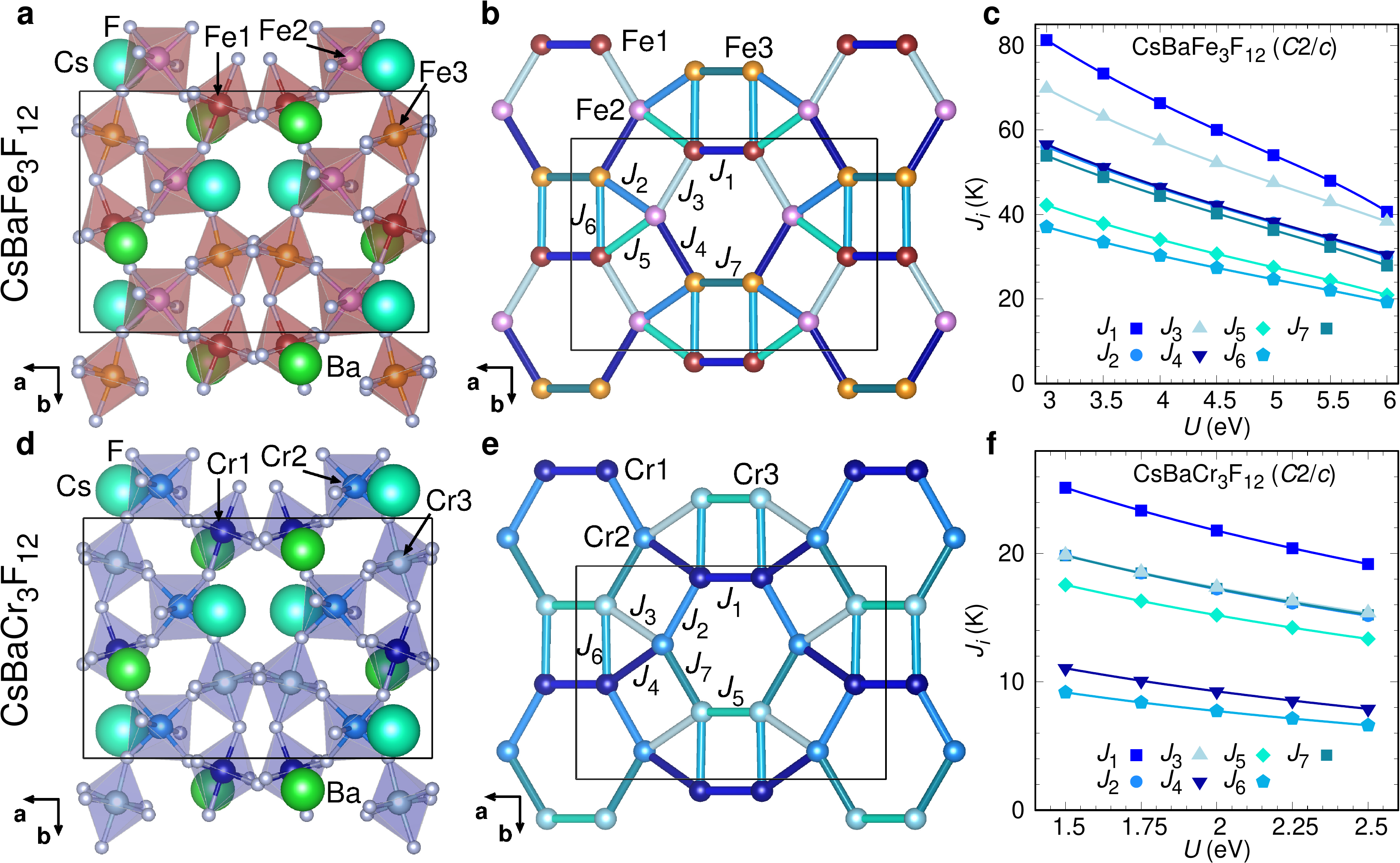}
\caption{{\bf Crystal structure, exchange paths and exchange interactions.} {\bf a}, {\bf d} Crystal structures of {\fe} and {\ch} viewed along the $c$ direction. {\bf b}, {\bf e} The first seven exchange paths that make up the ruby lattice of {\fe} and {\ch}. {\bf c}, {\bf f} Values of the seven nearest neighbour exchange interactions as a function of on-site interaction strength $U$ for {\fe} and {\ch}.}
\label{fig:structurecouplings}
\end{figure*}

The two compounds investigated in this work, {\fe} and {\ch}, were synthesized in a long-standing effort to discover novel fluoride compounds~\cite{Renaudin1986}, and their transition metal framework has been discussed in the context of mosaics and geometric tilings~\cite{Ferey1989,Ferey2014}. Both compounds are isostructural and crystallize in the monoclinic $C2/c$ space group; \ce{Fe3F12} or \ce{Cr3F12} form layers of corner-sharing \ce{FeF6} and \ce{CrF6} octahedra, respectively, forming ruby-lattice networks. These layers, shown in Fig.~\ref{fig:structurecouplings}a for {\fe} and Fig.~\ref{fig:structurecouplings}d for {\ch}, are separated by alkali and alkaline-earth ions located above and below the magnetic planes. A key structural feature is the pronounced quasi-two-dimensionality of these materials. The layer separations, $c \cdot \cos\beta=4.27$\,{\AA} for {\fe} and $c \cdot \cos\beta=4.24$\,{\AA} for {\ch}, substantially exceed the characteristic nearest-neighbour distances in the ruby lattice planes, making both materials excellent candidates for realizing intrinsic two-dimensional ruby-lattice magnetism.

We employ density functional theory (DFT) to establish that both compounds provide close realizations of the ruby-lattice Heisenberg antiferromagnet. The resulting effective spin models are then analyzed using Luttinger–Tisza (LT) theory and classical Monte Carlo (cMC) simulations to determine their magnetic properties and finite-temperature behavior. The magnetic ions carry spin $S=5/2$ in {\fe} and $S=3/2$ in {\ch}, placing both systems in the semiclassical regime. 
This motivates a combined approach based on LT theory to identify candidate ordering vectors and cMC simulations to incorporate thermal fluctuations and entropic effects.
Our analysis shows that both compounds realize an incommensurate spiral low-temperature state. For {\fe}, the theoretical results are in good agreement with available experimental data~\cite{Renaudin1991}, supporting the validity of the extracted effective model. For {\ch}, we find a particularly shallow energy landscape in momentum space around the ground state ordering vector, leading to a strong sensitivity of the ordering wave vector to thermal fluctuations, with cMC simulations finding temperature-dependent shifts in the ordering vector driven by entropic selection effects. To enable direct comparison with future neutron scattering experiments, we further compute the equal-time spin structure factor $S(\mathbf{q})$ for both compounds, providing clear signatures of the underlying incommensurate spiral correlations and their thermal evolution.

\begin{table*}[t]
\caption{{\bf Exchange paths and Heisenberg Hamiltonian couplings as determined from DFT-based energy mapping.} The paths are identified by Fe--Fe and Cr--Cr distance, respectively, and they are labelled according to increasing distance. {\it hex} and {\it tri} refer to hexagon and triangle, respectively. The arrangement of Cr1, Cr2, and Cr3 is the same as that of Fe1, Fe2, and Fe3, but their bond lengths are ordered differently. Fe couplings are for $S=5/2$, Cr couplings for $S=3/2$. 
}
\label{tab:couplings}
\centering
\vspace{1mm}
\begin{minipage}{0.48\textwidth}
\centering
{\fe}

\renewcommand{\arraystretch}{1.25}
\setlength{\tabcolsep}{7pt}

\begin{tabular}{c|c|l|c|c}
\hline
name & $d_{\rm Fe-Fe}$\,(\AA) & role & $J_i$\,(K) & $J_i/J_1$\\
\hline
$J_1$ & 3.62096 & Fe1-Fe1 hex & 47.98(3) & 1\\
$J_2$ & 3.64416 & Fe2-Fe3 tri & 34.17(3) & 0.712\\
$J_3$ & 3.65084 & Fe1-Fe2 hex & 42.93(3) & 0.895\\
$J_4$ & 3.71377 & Fe2-Fe3 hex & 34.47(3) & 0.718\\
$J_5$ & 3.72851 & Fe1-Fe2 tri & 24.42(3) & 0.509\\
$J_6$ & 3.74459 & Fe1-Fe3 tri & 22.04(3) & 0.459\\
$J_7$ & 3.7517  & Fe3-Fe3 hex & 32.33(3) & 0.674\\
\hline
\end{tabular}
\end{minipage}
\hfill
\begin{minipage}{0.48\textwidth}
\centering
{\ch}
\renewcommand{\arraystretch}{1.25}
\setlength{\tabcolsep}{7pt}
\begin{tabular}{c|c|l|c|c}
\hline
name & $d_{\rm Cr-Cr}$\,(\AA) & role & $J_i$\,(K) & $J_i/J_1$\\
\hline
$J_1$ & 3.62588 & Cr1-Cr1 hex & 21.80(4) & 1\\
$J_2$ & 3.62758 & Cr1-Cr2 hex & 17.23(2) & 0.790\\
$J_3$ & 3.63513 & Cr2-Cr3 tri & 17.35(2) & 0.796\\
$J_4$ & 3.64114 & Cr1-Cr2 tri & 9.27(3)  & 0.425\\
$J_5$ & 3.64615 & Cr3-Cr3 hex & 15.21(1) & 0.698\\
$J_6$ & 3.65495 & Cr1-Cr3 tri & 7.73(1)  & 0.354\\
$J_7$ & 3.65838 & Cr2-Cr3 hex & 17.27(2) & 0.792\\
\hline
\end{tabular}
\end{minipage}

\end{table*}

\section{Density Functional Theory Results}
We extract the Heisenberg Hamiltonian of {\fe} and {\ch} using DFT+$U$~\cite{Liechtenstein1995} energy mapping, which has yielded excellent results for various Cr based magnets~\cite{Ghosh2019,Xu2023,Nilsen2025}. We perform our calculations for the experimental crystal structures given in Ref.~\onlinecite{Renaudin1991} for {\fe} and in Ref.~\onlinecite{Ferey1989} for {\ch}. More details are given in section \ref{app:dft}.

Figs.~\ref{fig:structurecouplings}c and f illustrate the result for the dominant Heisenberg exchange interactions for {\fe} and {\ch}, respectively, as a function of the on-site interaction $U$. In both cases, we find that there are seven dominant interactions making up the nearest-neighbour (NN) bonds of the corresponding ruby lattices. Longer range interactions, including interlayer interactions, are found to be only about 1.5{\%} or less of the largest coupling $J_1$. This makes the Hamiltonian of both {\fe} and {\ch} very short-ranged, similar to other fluoride magnets~\cite{Jeschke2019,Shirakami2019}.

To pinpoint the precise value of the exchange interactions, our method requires some experimental input that characterizes the magnetic energy scale of the material; this could be a Curie-Weiss temperature, for example. As such information has not been determined for either of our two target materials, we proceed as follows: for {\fe}, we  select the Hamiltonian calculated at an on-site interaction value $U=5.5$\,eV because the feature connected to magnetic order in the calculated specific heat is close to the experimental Ne\'el temperature for this $U$ value. We have performed cMC calculations for six Hamiltonians obtained with different $U$ values and find that the behaviour does not change qualitatively (see {\suppSecMonteCarlo}). However, as the energy scale as quantified by the calculated Curie-Weiss temperature decreases from $\theta_{\rm CW}=-585$\,K at $U=3.5$\,eV to $\theta_{\rm CW}=-340$\,K at $U=6$\,eV, the features in the specific heat systematically shift to lower energies as $U$ is increased. Note that experiments have determined the magnetic susceptibility only up to $T=290$\,K~\cite{Renaudin1991}, which is insufficient to ascertain the energy scale of {\fe}.

For {\ch}, on the other hand, the magnetic properties have not been experimentally investigated so far. This means that the experimental Curie-Weiss temperature that we would normally use to fix the energy scale of the DFT energy mapping calculations is unknown. As a workaround, we use a value of $U=2$\,eV which is very typical for insulating Cr$^{3+}$ based antiferromagnets~\cite{Ghosh2019,Jaubert2025}.

The resulting interactions are given in Table~\ref{tab:couplings}. Noticeably, for both compounds, all nearest-neighbour (NN) couplings on the ruby lattice are antiferromagnetic. Apparently, there are subtle differences in the hierarchy of exchange interactions between the two compounds. What stands out for {\ch} are two relatively small triangle couplings; as they combine the unfrustrated hexagon couplings into a frustrated 2D lattice, the Hamiltonian of {\ch} at first glance appears somewhat less frustrated than that of {\fe}.

\section{Classical Monte Carlo Results}
Building on the effective spin Hamiltonians derived from DFT, we now investigate their magnetic properties using classical Monte Carlo (cMC) simulations.
We consider the Heisenberg model,
\begin{equation}
    H = \sum_{i<j} J_{ij}\, \mathbf S_i \cdot \mathbf S_j \,,
\label{eq:hamiltonian}
\end{equation}
of {\fe} and {\ch}. Technical details of the simulations are provided in Appendix \ref{app:morecmc}. In the following, we present the results for the two compounds separately.

\begin{figure}[t]
\includegraphics*[width=0.95\columnwidth]{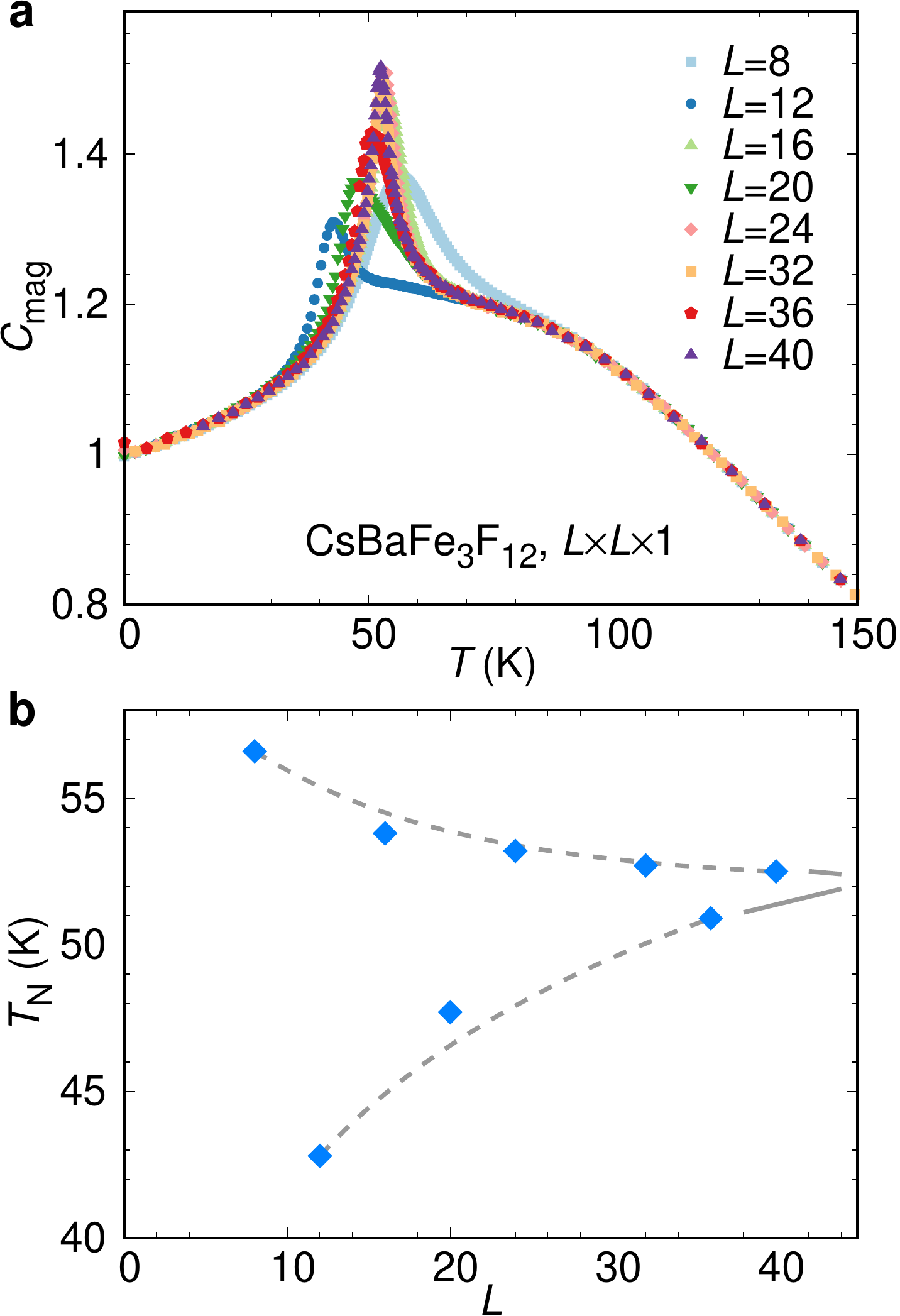}
    \caption{\textbf{Specific heat and magnetic ordering temperatures of {\fe} as a function of simulated lattice size.} {\bf a} Specific heat of {\fe} calculated for different system sizes. {\bf b} N{\'e}el temperatures extracted from maxima of specific heat in {\bf a}. The dashed lines are guides to the eye.}
\label{fig:specificheat_fe}
\end{figure}

\begin{figure}[t]
\includegraphics[width=\linewidth]{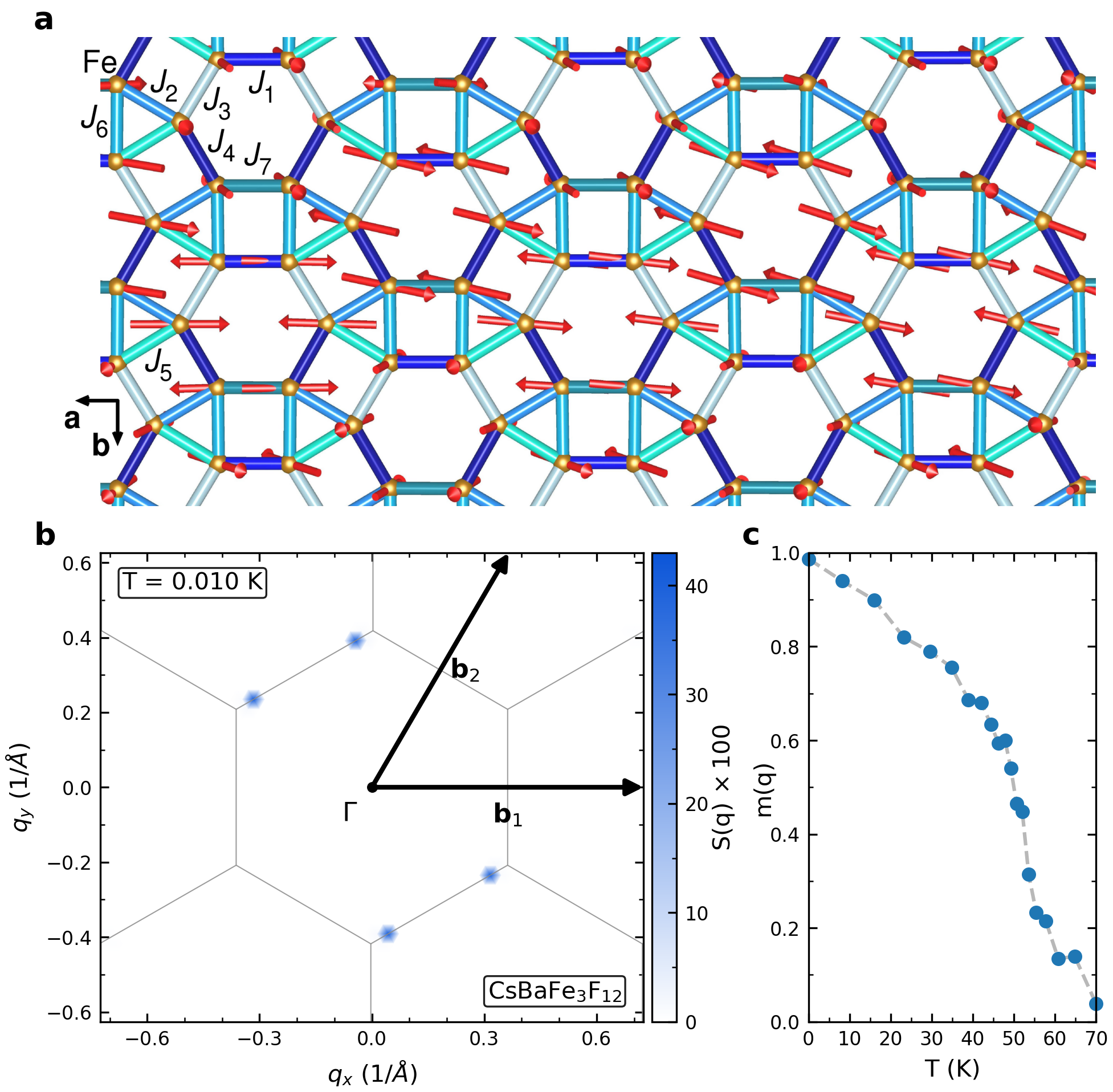}
\caption{\textbf{Ground state spin configuration, Equal-time spin structure factor and magnetic order parameter of {\fe}.} {\bf a} Real space spin configuration at T=0.01~K (proxy for the ground state). {\bf b} Equal-time spin structure factor, $S(\mathbf q)$, at T=0.01~K. The data are shown in the Cartesian $q_x$-$q_y$ plane. The central hexagon represents the first Brillouin zone (BZ). The calculation used a $24 \times 24$ supercell and a $48 \times 48$ grid for the momentum ${\bf q}$. {\bf c} Magnetic order parameter calculated from the overlap of finite-temperature spin configurations with the magnetic ground state in {\bf a}.}
\label{fig:spin_config_fe}
\end{figure}

\subsection{{\fe}}
For {\fe}, we performed cMC simulations for linear system sizes $L \in {8, \ldots, 40}$ unit cells. The magnetic specific heat $C_{\mathrm{mag}}$ exhibits two characteristic features: a pronounced peak near $T \approx 50\,\mathrm{K}$ and a broader shoulder around $T \sim 80\,\mathrm{K}$. The position of the main peak shows a dependence on system size, as shown in Fig.~\ref{fig:specificheat_fe}a, and becomes progressively sharper with increasing $L$, consistent with the development of a thermodynamic transition.
The finite-size evolution of the peak position is summarized in Fig.~\ref{fig:specificheat_fe}b, showing a systematic variation with increasing system size. Extrapolating to the thermodynamic limit suggests convergence to $T \approx 52$~K. We therefore associate this feature with the onset of long-range magnetic order and identify it with the N\'eel temperature $T_{\mathrm{N}} = 55(2)\,\mathrm{K}$ determined experimentally from M\"ossbauer spectroscopy~\cite{Renaudin1991}.

The size-dependent shoulder in the magnetic specific heat $C_{\mathrm{mag}}$ around $T \sim 80\,\mathrm{K}$, see Fig.~\ref{fig:specificheat_fe}a, is indicative of a crossover associated with the development of short-range spin correlations. This is consistent with experimental indications of an onset of short-range order around $100\,\mathrm{K}$, in agreement with our findings. At higher temperatures, $C_{\mathrm{mag}}$ decreases smoothly toward the paramagnetic regime and exhibits no finite-size dependence.

The magnetic order obtained at the lowest temperature is found to be a spin-spiral state and is shown in Fig.~\ref{fig:spin_config_fe}a. Interestingly, similar to the magnetic order in the AFM Heisenberg model on the ideal ruby lattice (Fig.~\ref{fig:ruby}), in this ordered state, each $J_1$-$J_3$-$J_4$-$J_7$-$J_4$-$J_3$ hexagon also assumes a local N\'eel order, which by the effect of the triangular interactions shows a relative inter-hexagonal canting.
The difference, however, is that the canting no longer happens with a commensurate angle, resulting in the incommensurate spin-spiral state. Please take note that the magnetic order we observe is consistent with the peaks found in M\"ossbauer spectroscopy.

To obtain a direct momentum-space signature of the magnetic correlations and to enable comparison with future neutron-scattering experiments, we further compute the equal-time spin structure factor (SSF) 
\begin{equation}
  S(\mathbf{q}) =
  \frac{1}{N}
  \sum_{l,m}
  \left\langle
  \mathbf{S}_l\cdot\mathbf{S}_m
  \right\rangle
  \exp\left[-i \mathbf{q}\cdot(\mathbf{r}_l-\mathbf{r}_m)\right],
\label{eq:ssf}
\end{equation}
from the cMC simulations. Here, $l$ and $m$ enumerate the lattice sites positioned at $\mathbf{r}_l$ and $\mathbf{r}_m$, respectively. The calculated SSF is shown in Fig.~\ref{fig:spin_config_fe}b. At the lowest temperature, the system shows Bragg peaks at two edges of the first Brillouin zone (BZ) located at the wavevector $\mathbf{q}_\text{Fe}= 0.375 \cdot ( \mathbf{b}_1 + \mathbf{b}_2) - \mathbf{b}_1$. We have rotated the reciprocal lattice vectors by $\phi = -30.262342^\circ$, so that $\mathbf{b}_1 = (0.725, 0.0)$~\AA$^{-1}$ and  $\mathbf{b}_2 = (0.365371, 0.626201)$~\AA$^{-1}$. In Fig.~\ref{fig:spin_config_fe}c, we show the projection of finite-temperature spin configurations onto the ground state, where the latter is taken as the spin configuration obtained at the lowest simulated temperature. Both the Bragg peaks in the SSF and the overlap with the magnetic ground state start to evolve around $55$~K, which is also consistent with the reported N\'eel temperature~\cite{Renaudin1991}. The peak strength in the SSF is weaker than one would expect for a pure N{\'e}el order. The reduced intensity is due to the large inter-hexagon canting in the magnetic order: the angle between the local N\'eel axes of adjacent hexagons is close to $\pi/2$, see Fig.~\ref{fig:spin_config_fe}a. As a result, the corresponding Bragg peaks may be weak and difficult to identify in neutron-scattering experiments, requiring careful analysis to distinguish them from the noise.
 
\begin{figure}[t]
\includegraphics*[width=0.95\columnwidth]{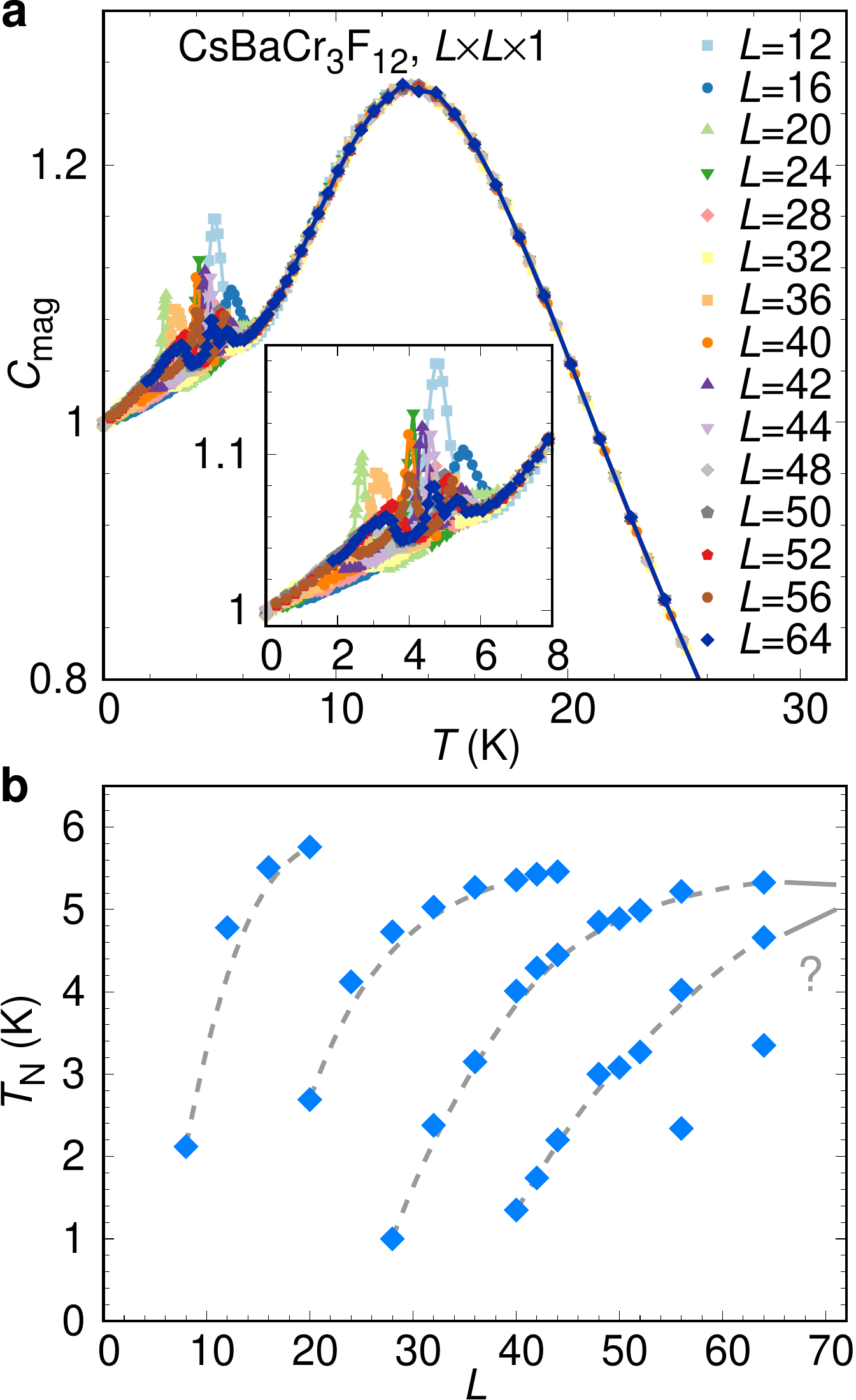}
\caption{\textbf{Specific heat and magnetic ordering temperatures of {\ch} as a function of simulated lattice size.} {\bf a} Specific heat of {\ch} calculated for different system sizes. {\bf b} N{\'e}el temperatures read off from peak maxima in {\bf a}. The dashed lines are a guide to the eye.}
\label{fig:specificheat_cr}
\end{figure}

\begin{figure*}[t]
\includegraphics[width=\linewidth]{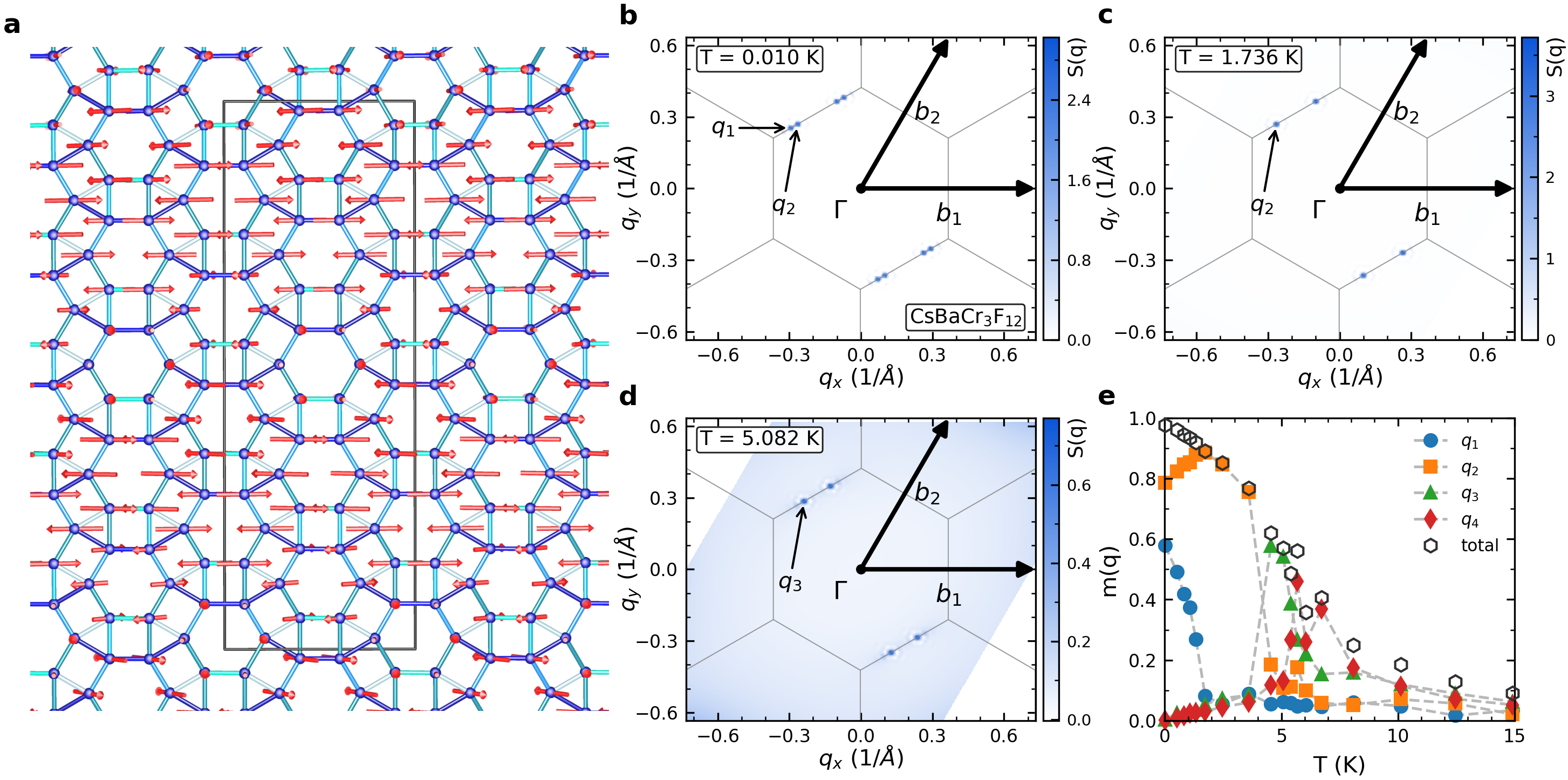}
\caption{\textbf{Ground state spin configuration, Equal-time spin structure factor and magnetic order parameter of {\ch}.} {\bf a} Real space spin configuration at T=0.01~K, approximating the ground state. {\bf b}-{\bf d} Evolution of the Equal-time spin structure factor of the ordered phase for $T\in \{0.01, 1.736, 5.082\}$~K. The data are shown in the Cartesian $q_x$-$q_y$ plane. The central hexagon represents the first Brillouin zone (BZ). The calculation used a $40 \times 40$ supercell and a $80 \times 80$ grid for the momentum $q$. {\bf e} Magnetic order parameter for the ordered states shown in panel {\bf b}-{\bf d}, calculated from the overlap of finite-temperature spin configurations with the respective magnetic states characterized by ordering vectors $q_1$, $q_2$, $q_3$. At higher temperatures, some fluctuations associated with $q_4$ appear, but do not lead to an ordered phase. The total order parameter is defined as \(m_{\mathrm{tot}} = \bigl(\sum_{i=1}^{4} m(q_i)^2\bigr)^{1/2}\).}
\label{fig:ssf_cr}
\end{figure*}

\subsection{{\ch}}
The magnetic specific heat $C_{\rm mag}$ for {\ch} is shown in Fig.~\ref{fig:specificheat_cr}a for a range of different linear system sizes $L$. For all system sizes, we observe a broad feature indicating the development of short-range correlations at about $T=13$\,K, see Fig.~\ref{fig:specificheat_cr}a. However, the features of $C_{\rm mag}$ at lower temperatures [see the inset of Fig.~\ref{fig:specificheat_cr}a] vary strongly with system size, showing one, two, or three not very sharp peaks. Moreover, the positions of these peaks shift significantly as a function of system size. The peak positions are summarized in Fig.~\ref{fig:specificheat_cr}b. This unusual behaviour of specific heat can be attributed to significant finite-size effects. Calculations for even larger systems are not possible, since the cMC runs become extremely costly at this point.
Based on the available results, we speculate that the peaks in the specific heat will eventually merge into one ordering peak approximately situated at $4$~K [please refer to Fig.~\ref{fig:specificheat_cr}b]. 

We think that the shifts and multiple low-temperature ordering peaks are caused mainly by the inability to accommodate an incommensurate order within a finite system size. In some cases, when the periodicity of such an incommensurate order is small, one can capture it approximately within a finite system. However, in our case, we find that this is not the case. 
In Fig.~\ref{fig:ssf_cr}a, we show the magnetic order assumed by the system at the lowest temperature ($0.001$~K) for a linear system size of $L=52$. Like {\fe} and the isotropic case discussed earlier, {\ch} also assumes a near-perfect local-N{\'e}el configuration on the $J_1$-$J_2$-$J_7$-$J_5$-$J_7$-$J_2$ hexagons with an inter-hexagon canting that appears because of the frustration introduced by the triangular interactions. In {\ch}, however, this canting induces a magnetic order, which is translationally invariant along the first lattice direction (Fig.~\ref{fig:structurecouplings}e), while the periodicity along its orthogonal direction is rather large. Further analysis of these short-range ordering tendencies is given in {\suppSecMonteCarlo}.

Our attempts to determine the exact periodicity of the order from cMC remain inconclusive, as we have not found a consistent periodicity with increasing system size. Most likely, this is a one-dimensional incommensurate magnetic order with a periodicity of approximately five unit cells. But, as we are constrained by finite system sizes, we cannot rule out the possibility that it can also be a commensurate order with a larger periodicity.

Now, we look at the maxima of the SSF as a function of temperature. A representative plot of the SSF for $L=40$ is shown in Fig.~\ref{fig:ssf_cr}\,(b-d). We have rotated the reciprocal lattice vectors by $\phi = -30.176140^\circ$, so that $\mathbf{b}_1 = (0.733236, 0.0)$~\AA$^{-1}$ and  $\mathbf{b}_2 = (0.368569, 0.633871)$~\AA$^{-1}$. From high to low temperature, the location of the maxima in the SSF shows discrete jumps before reaching the final ground state value. From high to low temperature the relevant wavevectors are $\mathbf{q}_4= 0.475 (\mathbf{b}_1 + \mathbf{b}_2) - \mathbf{b}_1$, $\mathbf{q}_3=0.45 (\mathbf{b}_1 + \mathbf{b}_2) - \mathbf{b}_1$ and $\mathbf{q}_2= 0.425 (\mathbf{b}_1 + \mathbf{b}_2) - \mathbf{b}_1$. Below $2$~K a further sub-dominant peak appears at $\mathbf{q}_1= 0.4 (\mathbf{b}_1 + \mathbf{b}_2) - \mathbf{b}_1$. The projection of spin configurations as a function of temperature onto the magnetic states associated with these wavevectors is shown in Fig.~\ref{fig:ssf_cr}e. For an overview, see Table~\ref{tab:reference-states}. For more details on the magnetic order parameters in this subfigure, see Appendix \ref{app:orderparameter} and {\suppSecWavevectors}. Our further analysis shows that the true ordering wavevector is probably located between $q_1$ and $q_2$.

\begin{table*}[t]
  \centering
  \renewcommand{\arraystretch}{1.25}
  \setlength{\tabcolsep}{10pt}
  \caption{Reference states for the overlap order
  parameters. The wave vectors are given in the rotated reciprocal space frame used in Figs.~\ref{fig:spin_config_fe} and \ref{fig:ssf_cr}. The order parameter \(m_{\rm ref}=m({\bf q}_\nu,T_{\rm ref})\) is maximized over all symmetry-equivalent momenta.}
  \label{tab:reference-states}
  \begin{tabular}{@{}lrcc@{}}
    \hline
    case
    & \multicolumn{1}{c}{\({\bf q}_\nu\)}
    & \(T_{\rm ref}\) (K)
    & \(m_{\rm ref}\)\\
    \hline
    Fe & \({\bf q}_{\rm Fe}= 0.375 (\mathbf{b}_1 + \mathbf{b}_2) - \mathbf{b}_1 = (-0.316,  0.235) \)\,\AA$^{-1}$ & 0.010 & 0.987  \\
    Cr & \({\bf q}_1= 0.400 (\mathbf{b}_1 + \mathbf{b}_2) - \mathbf{b}_1 = (-0.293,  0.254)\)\,\AA$^{-1}$ & 0.010 & 0.580  \\
    Cr & \({\bf q}_2= 0.425 (\mathbf{b}_1 + \mathbf{b}_2) - \mathbf{b}_1 = (-0.265,  0.269)\)\,\AA$^{-1}$ & 1.736 & 0.885  \\
    Cr & \({\bf q}_3= 0.450 (\mathbf{b}_1 + \mathbf{b}_2) - \mathbf{b}_1 = (-0.237,  0.285)\)\,\AA$^{-1}$ & 4.557 & 0.577  \\
    Cr & \({\bf q}_4= 0.475 (\mathbf{b}_1 + \mathbf{b}_2) - \mathbf{b}_1 = (-0.210,  0.301)\)\,\AA$^{-1}$ & 5.685 & 0.459  \\
    \hline
  \end{tabular}
\end{table*}

For our magnetic order with such a large unit-cell, a number of commensurate and, possibly, incommensurate orders should exist, which are in close competition with the true ground state of the system.
In frustrated magnets with shallow minima or extended manifolds of nearly degenerate spiral states, fluctuations can lift the accidental degeneracy and select particular ordering wave vectors through an order-by-disorder mechanism~\cite{Villain1980, Henley1989, Bergman2007, Mulder2010}. In the present finite-size cMC simulations this selection takes the form of discrete jumps of the dominant peak position in $S ({\bf q})$, because only a discrete set of momenta is compatible with a finite simulation supercell. We therefore interpret the observed sequence of nearby peak positions not as evidence for distinct, well-separated ordered phases, but as finite-size manifestations of a shallow incommensurate ordering minimum whose thermally selected wave vector lies between the accessible commensurate momenta.

\begin{figure}
\centering
\includegraphics[width=\linewidth]{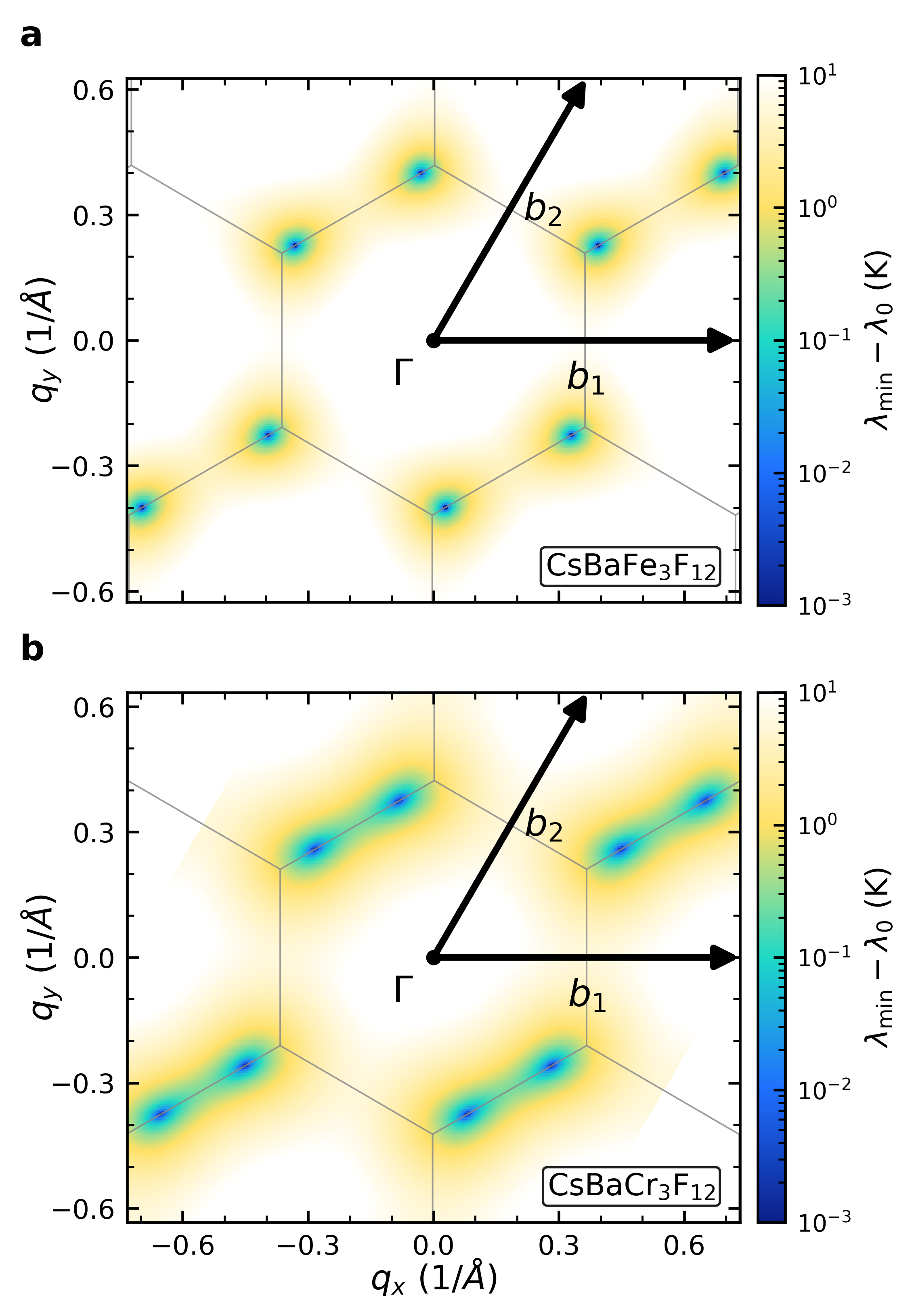}
\caption{{\bf Lowest eigenvalues of the dressed Luttinger-Tisza exchange matrix.} Panel {\bf a} shows the isolated minima for {\fe}. {\bf b} shows that the minima for {\ch} are connected by shallow plateaus. The central hexagon represents the first Brillouin zone.}
\label{fig:luttingertisza}
\end{figure}

\section{Luttinger-Tisza analysis}
Complementary to our cMC calculations, we analyze the full classical Heisenberg Hamiltonian (Eq.~\ref{eq:hamiltonian}) for {\fe} and {\ch} using an extension of the standard Luttinger-Tisza (LT) method suitable non-Bravais lattice systems~\cite{LT,Schmidt2022}. The LT method relaxes the hard-spin constraint and reduces the problem of finding the ground state ordering vector to minimizing the lowest eigenvalue of the Fourier-transformed exchange matrix $\mathcal{J}_{\alpha \beta} ({\bf q})$. This procedure yields a unique ${\bf {\hat q}}$, which is the ordering vector of a single-${\bf q}$ coplanar spiral. For more details, see Appendix \ref{app:luttinger}.

Note that we implement the LT method for a unit spin $S=1$, while {\fe} and {\ch} are $S=5/2$ and $S=3/2$, respectively. Therefore, all energy values should be multiplied by $S^2$ to obtain physical values. For non-Bravais systems there is no guarantee that the spin length constraint will be satisfied perfectly. For the compounds we investigate here, the hard-spin constraint can be substantially violated throughout the BZ, but this violation is not very drastic near the boundary of the BZ, and the hard-spin constraint is in fact perfectly fulfilled on the BZ boundary where the ordering wave vectors are located. 

The results of our LT calculations are summarized in Fig.~\ref{fig:luttingertisza}, where we show the lowest eigenvalue of the LT dressed exchange matrix in momentum space. For {\fe}, we find a classical ordering vector of $\hat {\bf q}_{\rm Fe} = 0.361611 (\mathbf{b}_1 + \mathbf{b}_2) - \mathbf{b}_1$ with a well-defined minimum of the lowest eigenvalue at $\lambda_\text{min} = -106.287$~K and a stiffness (curvature of the LT spectrum) of $\kappa_\text{Fe} = 169.938$~\AA$^2$K, see Fig.~\ref{fig:luttingertisza}a. The ordering vector is very close to ${\bf q}_{\rm Fe} = 0.375 (\mathbf{b}_1 + \mathbf{b}_2) - \mathbf{b}_1$ as determined from cMC (see Table~\ref{tab:reference-states}).

For {\ch}, we find a classical ordering vector of $\hat {\bf q}_{\rm Cr} = 0.408173 (\mathbf{b}_1 + \mathbf{b}_2) - \mathbf{b}_1$ with a shallow minimum of the eigenvalue around $\lambda_\text{min} = -48.446$~K with stiffness $\kappa_\text{Cr} = 39.050$~\AA$^2$K. The ordering vector is very close to ${\bf q}_1 = 0.4 (\mathbf{b}_1 + \mathbf{b}_2) - \mathbf{b}_1$ and ${\bf q}_2 = 0.425 (\mathbf{b}_1 + \mathbf{b}_2) - \mathbf{b}_1$ determined from cMC (see Table~\ref{tab:reference-states}). One notable feature of Fig.~\ref{fig:luttingertisza}b is that the two close-by minima in {\ch} are connected by a relatively shallow plateau, as indicated by the much lower stiffness compared to {\fe}. 

The origin of this stiffness contrast can be understood from an effective hexagon-spin description, derived in {\suppSecStiffnessModel}. In this projection, each locally N\'eel-ordered hexagon is represented by one staggered unit vector, and the full exchange problem reduces to an effective triangular network of hexagon N\'eel vectors. The resulting LT eigenvalue depends only on two projected triangular coupling sums, which control the location and stiffness of the ordering wavevector. If the strength of the triangular exchange couplings could be tuned selectively, e.g.~via anisotropic strain or pressure, the stiffness of the ordering vector could be accessed experimentally.

\section{Conclusions and Discussion}
In this work, we have investigated two transition metal fluorides, namely {\fe} and {\ch}, in which the the magnetic ions, i.e.~Fe$^{3+}$ and Cr$^{3+}$, assume slightly distorted ruby lattice structures with spin moments $S=5/2$ and $S=3/2$, respectively. 
Using DFT energy mapping, we extracted the corresponding Heisenberg Hamiltonians and established that both materials are governed predominantly by antiferromagnetic nearest-neighbour interactions.
While the exchange couplings on both the hexagonal and triangular bonds are antiferromagnetic and comparable in magnitude, the interactions along the hexagonal links are sufficiently stronger to place the systems in a regime of frustrated ruby-lattice magnetism.

Owing to the rather large spin moments of the magnetic ions, we analyzed the resulting spin models using classical Monte Carlo simulations together with Luttinger–Tisza theory to determine their low-temperature magnetic states and thermodynamic properties.
In both compounds, the dominant antiferromagnetic couplings on the hexagonal bonds promote a near-perfect N\'eel arrangement on the hexagons. The weaker competing interactions associated with the triangular motifs then cant these locally ordered hexagonal units relative to one another, ultimately stabilizing incommensurate spiral magnetic orders in both compounds at low temperatures.

Despite the comparable overall energy scales of the exchange interactions in the two compounds, their ordering temperatures obtained from the cMC calculations differ strikingly: {\fe} orders at approximately $50$~K, whereas {\ch} develops magnetic order only near $5$~K. A natural contributing factor is the larger spin length of the Fe$^{3+}$ ions, which enhances the stability of classical magnetic order. Beyond this, however, our analysis reveals an important qualitative distinction between the two systems. For {\fe}, the exchange couplings select a well-defined incommensurate spiral state with a comparatively robust ordering wave vector. In contrast, our LT analysis shows that {\ch} hosts an unusually shallow energy landscape in momentum space. In this case, the frustrating triangular interactions generate an extended manifold of nearly degenerate spiral configurations with very similar energies. The resulting proliferation of low-lying classical states strongly enhances fluctuation effects and suppresses the buildup of long-range magnetic order, thereby driving the ordering temperature to substantially lower values. If we calculate the stiffness ratio and correct the energy scale by the physical spin length, we obtain an estimate for the ordering temperature ratio in both compounds:
\begin{equation}
\frac{\kappa_\text{Fe}}{\kappa_\text{Cr}} \frac{S^2_\text{Fe}}{S^2_\text{Cr}} = \frac{169.938\,\text{\AA}^2\mathrm{K}}{39.050\,\text{\AA}^2\mathrm{K}} \cdot \frac{(5/2)^2}{(3/2)^2} \approx 12.088 \,.
\end{equation}
Therefore, we expect the ordering temperature in {\ch} to be about 12 times lower than in {\fe}, i.e.~around $4.5$~K. This is consistent with the ordering scale observed in our cMC simulations of {\ch}.

Another consequence of the unusually shallow minimum is that the energy landscape provides only weak selection of a specific $q$, with many nearby $\mathbf{q}$ points possessing almost identical energies. In such a situation, the ordering vector is not rigidly fixed by energetics alone.
At finite temperature, the relevant quantity is the free energy. Different points along this quasi-degenerate manifold generally support different fluctuation spectra. Some ordering vectors sit in flatter regions of the dispersion, for example along the boundary of the BZ, and therefore allow a larger phase space of low-energy excitations, resulting in an enhanced entropic contribution. As temperature increases, these entropic effects become comparable to the minute energy differences within the shallow minimum, shifting the balance between nearly equivalent wavevectors $\bf q$. Consequently, the system reorganizes its ordering vector with temperature. This explains the temperature-dependent shift of the ordering vector observed in our cMC simulations of the {\ch} Hamiltonian as an entropic selection effect operating within a shallow manifold of low-lying spiral states. 

The near-perfect N\'eel orders in the two compounds also signal that the systems are very close to the ideal ruby limit. However, it also indicates that a slight asymmetry between the interactions can change the details of the magnetic order in a non-trivial way resulting in incommensurate spin spiral states with shallow energy landscapes. This is a strong motivation to look for further members of the \ce{CsBa$T$3F12} ruby lattice family of compounds by synthesizing variants with different transition metal ions $T^{3+}$ carrying smaller magnetic moments, where quantum fluctuations are expected to play a prominent role. Considering that corner sharing kagome networks of \ce{VF6} and \ce{TiF6} octahedra have been realized in chemically very similar compounds like \ce{Cs2KV3F12}~\cite{Goto2017} or \ce{Cs2KTi3F12}~\cite{Goto2016}, it appears plausible that ruby lattice fluorides \ce{CsBaV3F12} or \ce{CsBaTi3F12} could exist; they would be extremely interesting as V$^{3+}$ $S=1$ and Ti$^{3+}$ $S=1/2$ ions would introduce quantum fluctuations into new probably antiferromagnetic ruby lattice realizations.

\section*{Acknowledgments}
P.G.~thanks Fr\'ed\'eric Mila for useful discussions.
H.O.J.~acknowledges support through JSPS KAKENHI Grants No.~24H01668 and No.~25K08460. Part of the computation in this work has been done using the facilities of the Supercomputer Center, the Institute for Solid State Physics, the University of Tokyo.
D.G.~acknowledges the support of the Investment Bank of the State of Brandenburg (ILB) under Grant No.~86001926, which was co-financed by the European Union through its regional development fund (ERDF).
P.G.~acknowledges financial support from the Swiss National Funds.   

\bibliographystyle{naturemag}
\bibliography{bounce}

\appendix
\section{DFT energy mapping}\label{app:dft}
We employ the full potential local orbital (FPLO) basis~\cite{Koepernik1999} in combination with a generalized gradient exchange correlation functional~\cite{Perdew1996}. We treat the strong electronic correlations on the $3d$ orbitals of Fe$^{3+}$ and Cr$^{3+}$ ions using DFT+$U$~\cite{Liechtenstein1995}, where we fix the Hund's rule coupling to the values proposed in Ref.~\onlinecite{Mizokawa1996}, $J_{\rm H}=0.8$\,eV for Fe and $J_{\rm H}=0.72$\,eV for Cr (see also Refs.~\onlinecite{Ghosh2019,Xu2023,Nilsen2025}). Our choices of $U = 5.5$~eV for {\fe} and $U = 2$~eV for {\ch} for the on-site interaction are detailed in the main text.

For both {\fe} and {\ch} we reduce the symmetry from $C2/c$ to $P1$, which gives us twelve inequivalent spins. We calculate energies of 50 out of 560 distinct spin configurations and fit them to the Heisenberg Hamiltonian in Eq.~\ref{eq:hamiltonian}. This allows us to extract 18 exchange interactions, which excellently reproduce the DFT energies. For further details, see {\suppSecEnergyMapping}.

\section{Classical Monte Carlo}\label{app:morecmc}
We performed cMC simulations for the Hamiltonian in Eq.~\ref{eq:hamiltonian} using the Metropolis-Hastings algorithm~\cite{Metropolis1953, Hastings1970} with local updates. The random spins for local updates are chosen according to a cone-based adaptive rule~\cite{AlzateCardona2019}, which keeps the acceptance rate of the Metropolis algorithm close to 50\%. After each Metropolis lattice sweep, we perform one energy-conserving overrelaxation sweep~\cite{Kadena1994} for the entire lattice. We use a supercell of size $L \times L \times 1$. Since the ruby lattice has 12 basis sites, this results in $12 L^2$ Fe/Cr sites in total.

The Heisenberg couplings $J_{ij}$ include all couplings determined from DFT, including small interlayer couplings, which go beyond those in Table~\ref{tab:couplings}. A complete overview of all couplings including long-range and interlayer interactions as well as the system size dependence of results and the simulation settings used in each case are given in {\suppSecMonteCarlo}.

\section{Luttinger-Tisza method}\label{app:luttinger}
The LT calculation is performed in the full 12-site magnetic basis of the primitive cell. For each wave vector \({\bf q}\), we Fourier transform the exchange Hamiltonian to obtain the \(12\times12\) Hermitian matrix \(\mathcal J_{\alpha\beta}({\bf q})\), where \(\alpha,\beta=1,\ldots,12\) label magnetic basis sites. The ordinary LT procedure minimizes the lowest eigenvalue of this matrix, but on a non-Bravais lattice the corresponding eigenvector need not satisfy the equal-spin-length constraint on all basis sites. We therefore use a dressed LT construction in which diagonal Lagrange shifts $\kappa_g$ are added to the exchange matrix,
\begin{equation}
\tilde{\mathcal J}_{\alpha\beta}({\bf q},\kappa)
=
\mathcal J_{\alpha\beta}({\bf q})
+
\kappa_{g(\alpha)}\delta_{\alpha\beta},
\qquad
\sum_g n_g\kappa_g=0 .
\end{equation}
Here \(g(\alpha)=1,2,3\) assigns each of the 12 magnetic basis sites to one of the three crystallographic transition-metal site groups, each containing \(n_g=4\) symmetry-related sites. The final constraint removes the arbitrary overall shift of all eigenvalues. For each \({\bf q}\), we maximize the lowest eigenvalue with respect to the three group shifts \(\kappa_g\), and then minimize the resulting dressed lowest eigenvalue with respect to \({\bf q}\). The resulting eigenstate has equal amplitudes on all magnetic basis sites for the ordering vectors reported here, and therefore gives a valid single-\({\bf q}\) coplanar spiral satisfying the hard-spin constraint.

\section{Definition of magnetic order parameters}\label{app:orderparameter}
Our definition of the magnetic order parameters is based on our analysis of the low-temperature spin configurations. We first use Fourier analysis on the simulated magnetic states to find a reference state. Subsequently, we calculate the overlap of simulated magnetic configurations with these reference states as a function of temperature. 

The total order parameter in Fig.~\ref{fig:ssf_cr}e combines the four leading low-temperature components by a root-sum-square:
\begin{equation}
  m_{\rm tot}(T)
  =
  \left[
    \sum_{\nu=1}^{4}
    m({\bf q}_\nu,T)^2
  \right]^{1/2}.
  \label{eq:total-order-parameter}
\end{equation}
For more details on these calculations and a detailed characterization of the magnetic configurations related to each ordering wavevector, see {\suppSecWavevectors}.

\clearpage


\section*{Supplementary material}

# Realization of the Ruby Lattice Antiferromagnet in Layered Transition-Metal Fluorides –Supplementary Information–

Harald O. Jeschke,[1] Daniel Guterding,[2] and Pratyay Ghosh[3]

[1] Research Institute for Interdisciplinary Science, Okayama University, Okayama 700-8530, Japan
[2] Technische Hochschule Brandenburg, Magdeburger Straße 50, 14770 Brandenburg an der Havel, Germany
[3] Institute of Physics, Ecole Polytechnique Fédérale de Lausanne (EPFL), CH-1015 Lausanne, Switzerland

## Supplementary Note 1. ADDITIONAL DETAILS OF THE DFT+$U$ ENERGY MAPPING

In Tables 1 and 2, we present all Hamiltonian parameters we have calculated for $CsBaFe_3F_{12}$ and $CsBaCr_3F_{12}$, respectively. The $U$ values we have chosen to focus on in this work are marked in bold face, and detailed justification for this choice is given in the main text.

In Fig. 1 we show the fit of our Heisenberg model w.r.t. the magnetic DFT+$U$ energies at $U = 5.5\,\mathrm{eV}$. The fit error is negligible.

## Supplementary Note 2. ADDITIONAL DETAILS ON THE CLASSICAL MONTE CARLO CALCULATIONS

For each simulation we perform 5000 lattice sweeps for the warmup. Subsequently we perform $N$ measurements, where for each measurement we first execute one Metropolis sweep across the lattice, followed by an energy-conserving overrelaxation sweep of the full lattice. In the case of magnetic susceptibilities and cMC total energies, we have chosen $N$ so that the resulting observables are well converged. The number of measurements for each compound and lattice size is documented in Table 3. In addition, each cMC run

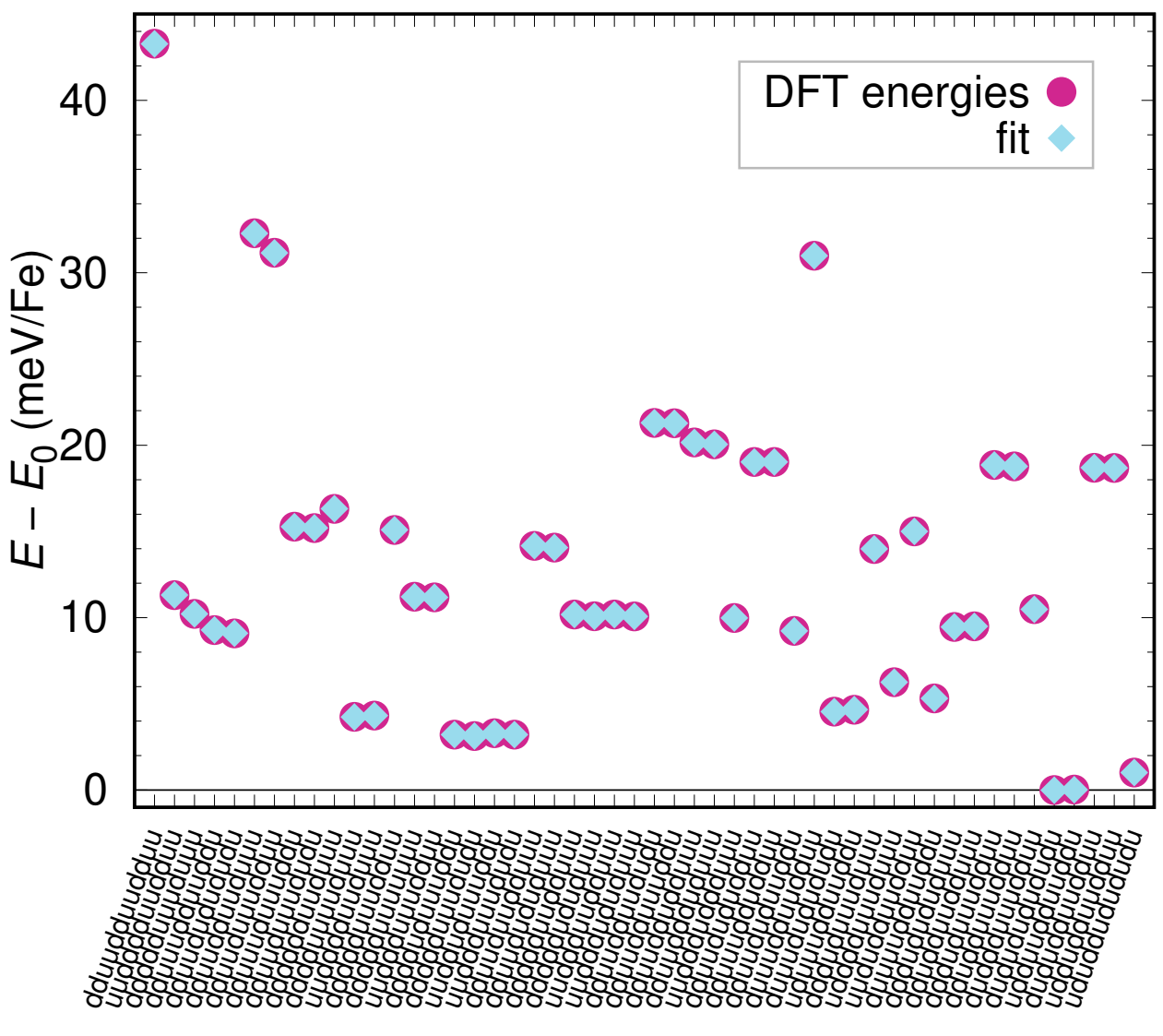


Supplementary Figure 1. DFT+$U$ energies for 50 spin configurations of $CsBaFe_3F_{12}$ at $U = 5.5\,\mathrm{eV}$, together with the fitted energies of the Heisenberg Hamiltonian.

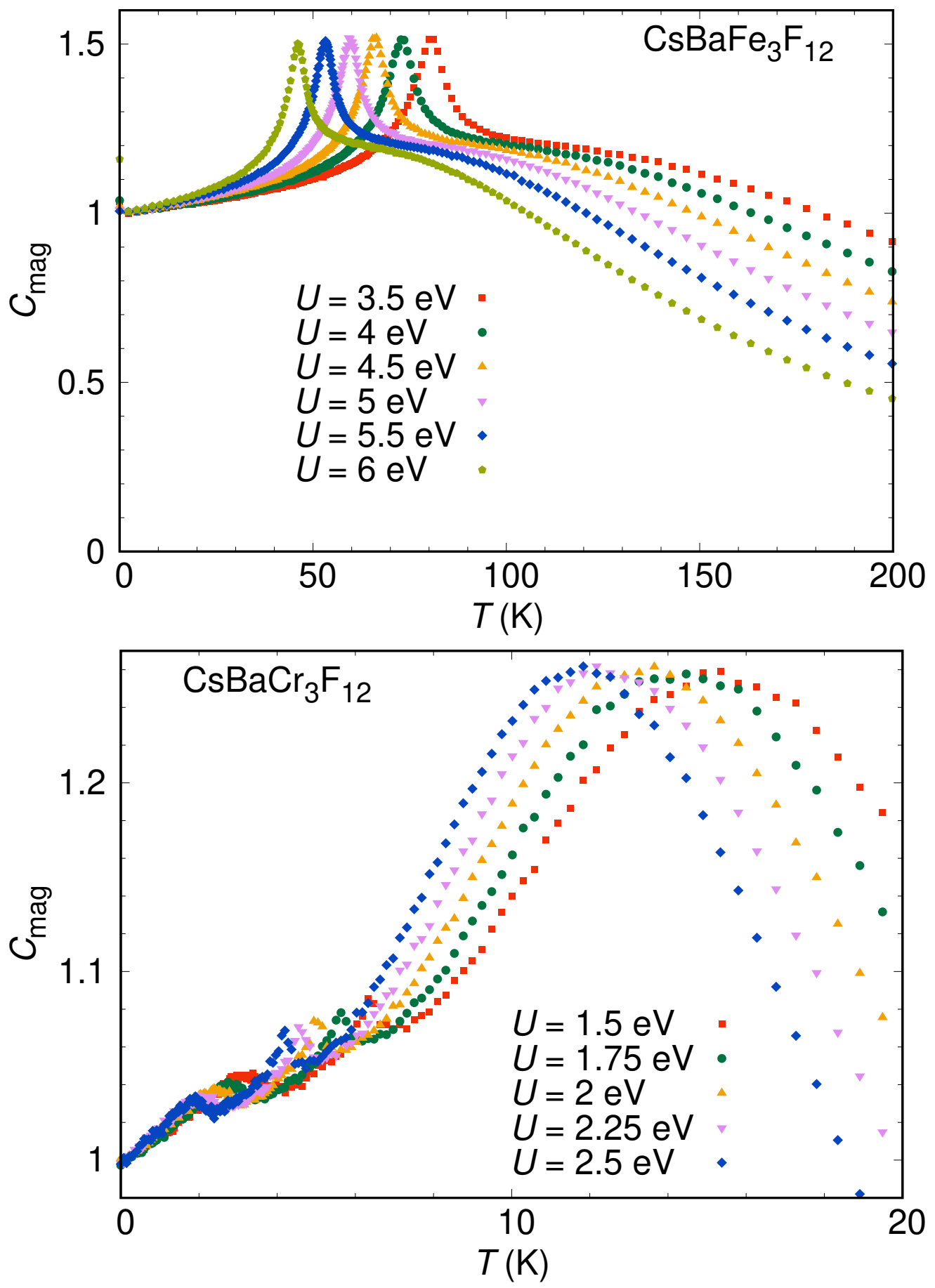


Supplementary Figure 2. Specific heat of $CsBaFe_3F_{12}$ (top) and $CsBaCr_3F_{12}$ (bottom) for Hamiltonians calculated with several different values of the on-site interaction strength $U$.

is averaged over the measurements of 512 parallel threads.

The bond angle histogram (see below) and SSF calculations deviate from this scheme. The bond angle histogram and SSF calculations for $CsBaFe_3F_{12}$ use a $L = 24$ supercell and $N = 60000$ measurements. For $CsBaCr_3F_{12}$ we use a $L = 40$ supercell and $N = 60000$ measurements. These results were averaged over 96 parallel threads.

In Fig. 2 we show how the energy scales visible in the magnetic specific heat change as a function of the on-site Coulomb interaction $U$.

We investigate the system size dependence of our cMC calculations via total energy per site (Fig. 3) and via specific heat (Fig. 4 in the main text). From $8 \times 8 \times 1$ supercells

Supplementary Table 1. Exchange couplings of $CsBaFe_3F_{12}$ calculated with DFT+$U$ energy mapping. The set of couplings we focus on here is given in bold face.

| $U$ (eV) | $J_1$ (K) | $J_2$ (K) | $J_3$ (K) | $J_4$ (K) | $J_5$ (K) | $J_6$ (K) | $J_7$ (K) | $\theta_{CW}$ (K) |
|---|---|---|---|---|---|---|---|---|
| 3 | 81.29(6) | 56.05(6) | 69.76(7) | 56.60(7) | 42.22(6) | 37.06(7) | 53.92(7) | -647.7 |
| 3.5 | 73.35(5) | 50.82(5) | 63.12(6) | 51.23(6) | 37.86(5) | 33.45(6) | 48.88(6) | -585.2 |
| 4 | 66.35(4) | 46.18(4) | 57.33(5) | 46.50(5) | 34.08(4) | 30.27(5) | 44.38(5) | -530.2 |
| 4.5 | 60.02(4) | 41.98(3) | 52.17(4) | 42.25(4) | 30.68(4) | 27.38(4) | 40.25(4) | -480.7 |
| 5 | 54.03(3) | 38.03(3) | 47.44(4) | 38.30(4) | 27.52(3) | 24.68(4) | 36.31(4) | -434.4 |
| **5.5** | **47.98(3)** | **34.17(3)** | **42.93(3)** | **34.47(3)** | **24.42(3)** | **22.04(3)** | **32.33(3)** | **-389.1** |
| 6 | 40.67(2) | 30.08(2) | 38.31(3) | 30.44(3) | 20.95(3) | 19.28(3) | 27.93(3) | -340.1 |
| $d_{Fe-Fe}$ (Å) | 3.621 | 3.644 | 3.651 | 3.714 | 3.729 | 3.745 | 3.752 | |

| $U$ (eV) | $J_8$ (K) | $J_9$ (K) | $J_{11}$ (K) | $J_{12}$ (K) | $J_{14}$ (K) | $J_{15}$ (K) | $J_{16}$ (K) | $J_{19}$ (K) | $J_{20}$ (K) | $J_{28}$ (K) | $J_{32}$ (K) |
|---|---|---|---|---|---|---|---|---|---|---|---|
| 3 | 0.40(5) | 0.69(6) | 0.25(6) | 0.94(7) | 0.10(4) | 0.15(7) | 1.13(7) | 0.34(7) | 0.04(7) | 0.06(6) | 0.01(3) |
| 3.5 | 0.34(4) | 0.60(5) | 0.21(5) | 0.84(5) | 0.09(4) | 0.13(6) | 1.02(6) | 0.30(5) | 0.03(6) | 0.05(5) | 0.01(3) |
| 4 | 0.29(3) | 0.51(4) | 0.18(4) | 0.75(5) | 0.08(3) | 0.11(5) | 0.93(5) | 0.27(5) | 0.02(5) | 0.05(4) | 0.01(2) |
| 4.5 | 0.24(3) | 0.44(4) | 0.16(4) | 0.68(4) | 0.07(3) | 0.10(4) | 0.83(4) | 0.23(4) | 0.02(4) | 0.04(4) | 0.01(2) |
| 5 | 0.20(2) | 0.37(3) | 0.13(3) | 0.60(3) | 0.06(2) | 0.09(4) | 0.74(3) | 0.19(3) | 0.01(3) | 0.04(3) | 0.01(2) |
| **5.5** | **0.16(2)** | **0.30(3)** | **0.11(3)** | **0.52(3)** | **0.06(2)** | **0.08(3)** | **0.64(3)** | **0.15(3)** | **0.01(3)** | **0.04(3)** | **0.01(2)** |
| 6 | 0.12(2) | 0.21(3) | 0.08(2) | 0.42(3) | 0.05(2) | 0.07(3) | 0.52(3) | 0.09(3) | 0.01(3) | 0.04(3) | 0.01(1) |
| $d_{Fe-Fe}$ (Å) | 5.197 | 5.222 | 5.943 | 6.150 | 6.272 | 6.272 | 6.273 | 6.633 | 6.689 | 7.160 | 7.547 |

Supplementary Table 2. Exchange couplings of $CsBaCr_3F_{12}$ calculated with DFT+$U$ energy mapping. The set of couplings we focus on here is given in bold face.

| $U$ (eV) | $J_1$ (K) | $J_2$ (K) | $J_3$ (K) | $J_4$ (K) | $J_5$ (K) | $J_6$ (K) | $J_7$ (K) | $\theta_{CW}$ (K) |
|---|---|---|---|---|---|---|---|---|
| 1.5 | 25.14(4) | 19.90(3) | 19.87(2) | 11.07(3) | 17.57(2) | 9.18(2) | 19.85(2) | -84.38 |
| 1.75 | 23.36(4) | 18.47(3) | 18.53(2) | 10.09(3) | 16.31(2) | 8.40(1) | 18.48(2) | -78.14 |
| **2** | **21.80(4)** | **17.23(2)** | **17.35(2)** | **9.27(3)** | **15.21(1)** | **7.73(1)** | **17.27(2)** | **-72.71** |
| 2.25 | 20.42(4) | 16.13(2) | 16.30(2) | 8.54(3) | 14.23(1) | 7.14(1) | 16.20(2) | -67.92 |
| 2.5 | 19.19(4) | 15.16(2) | 15.36(2) | 7.90(3) | 13.35(1) | 6.62(1) | 15.23(2) | -63.62 |
| $d_{Cr-Cr}$ (Å) | 3.626 | 3.628 | 3.635 | 3.641 | 3.646 | 3.655 | 3.658 | |

| $U$ (eV) | $J_8$ (K) | $J_9$ (K) | $J_{11}$ (K) | $J_{13}$ (K) | $J_{16}$ (K) | $J_{17}$ (K) | $J_{18}$ (K) | $J_{19}$ (K) | $J_{20}$ (K) | $J_{29}$ (K) | $J_{32}$ (K) |
|---|---|---|---|---|---|---|---|---|---|---|---|
| 1.5 | 0.11(2) | 0.20(2) | -0.03(3) | -0.17(2) | -0.14(2) | 0.00(2) | 0.03(2) | 0.00(2) | 0.01(2) | 0.01(2) | -0.01(2) |
| 1.75 | 0.09(1) | 0.16(2) | -0.03(3) | -0.16(2) | -0.13(2) | 0.00(2) | 0.03(2) | 0.01(2) | 0.01(2) | 0.01(2) | -0.01(2) |
| **2** | **0.06(1)** | **0.11(1)** | **-0.03(3)** | **-0.15(2)** | **-0.13(2)** | **0.00(1)** | **0.02(2)** | **0.01(2)** | **0.01(2)** | **0.01(1)** | **-0.01(2)** |
| 2.25 | 0.04(1) | 0.07(1) | -0.03(3) | -0.15(2) | -0.12(2) | 0.00(1) | 0.02(2) | 0.01(2) | 0.01(2) | 0.01(1) | -0.01(2) |
| 2.5 | 0.02(1) | 0.03(1) | -0.04(3) | -0.14(2) | -0.12(2) | 0.00(1) | 0.02(2) | 0.01(2) | 0.01(2) | 0.01(1) | -0.01(2) |
| $d_{Cr-Cr}$ (Å) | 5.134 | 5.153 | 6.051 | 6.270 | 6.309 | 6.384 | 6.406 | 6.776 | 6.867 | 7.139 | 7.593 |

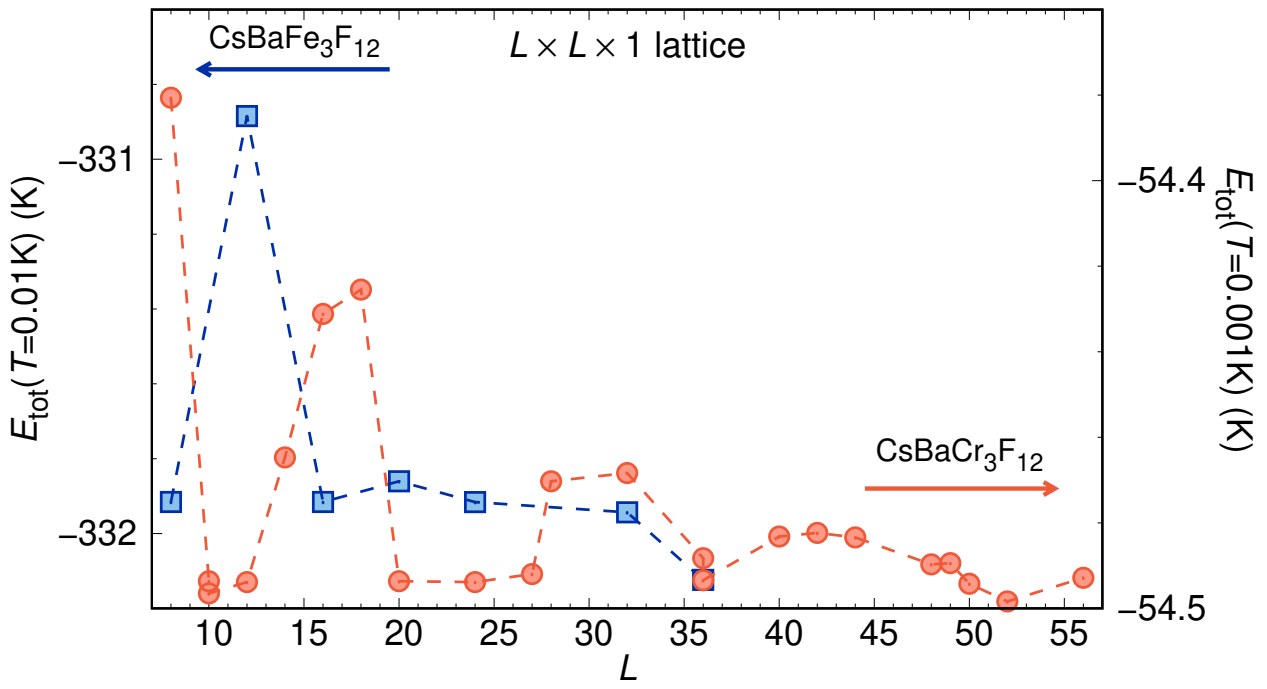


Supplementary Figure 3. Total energy per site for $CsBaFe_3F_{12}$ (left axis) and $CsBaCr_3F_{12}$ (right axis).

(768 spins) or larger, the overall features of the specific heat, namely a broad short-range-correlation maximum and an ordering-related peak at lower temperature, are present.

For $CsBaFe_3F_{12}$, the position in temperature of the ordering feature oscillates somewhat with increasing system size. For $CsBaCr_3F_{12}$, up to three ordering features appear but some of the system sizes show only the lower or the higher temperature peak. Many measurements for large system sizes are required to properly resolve the rather narrow ordering features.

We also calculated the histogram of bond angles for each exchange interaction $J_i$ as a function of temperature with cMC. For $CsBaFe_3F_{12}$ the bonds $J_1$, $J_3$, $J_4$ and $J_7$ are hexagon bonds, while $J_2$, $J_5$ and $J_6$ are triangular bonds. The data in Fig. 4 shows that correlations along the hexagon bonds start at much higher temperatures than for the triangular bonds, supporting the notion that $CsBaFe_3F_{12}$ shows local-Neél configurations on the hexagons above the ordering temperature.

We observe a similar behaviour for $CsBaCr_3F_{12}$, where $J_1$, $J_2$, $J_5$ and $J_7$ are the hexagon bonds, while $J_3$, $J_4$ and

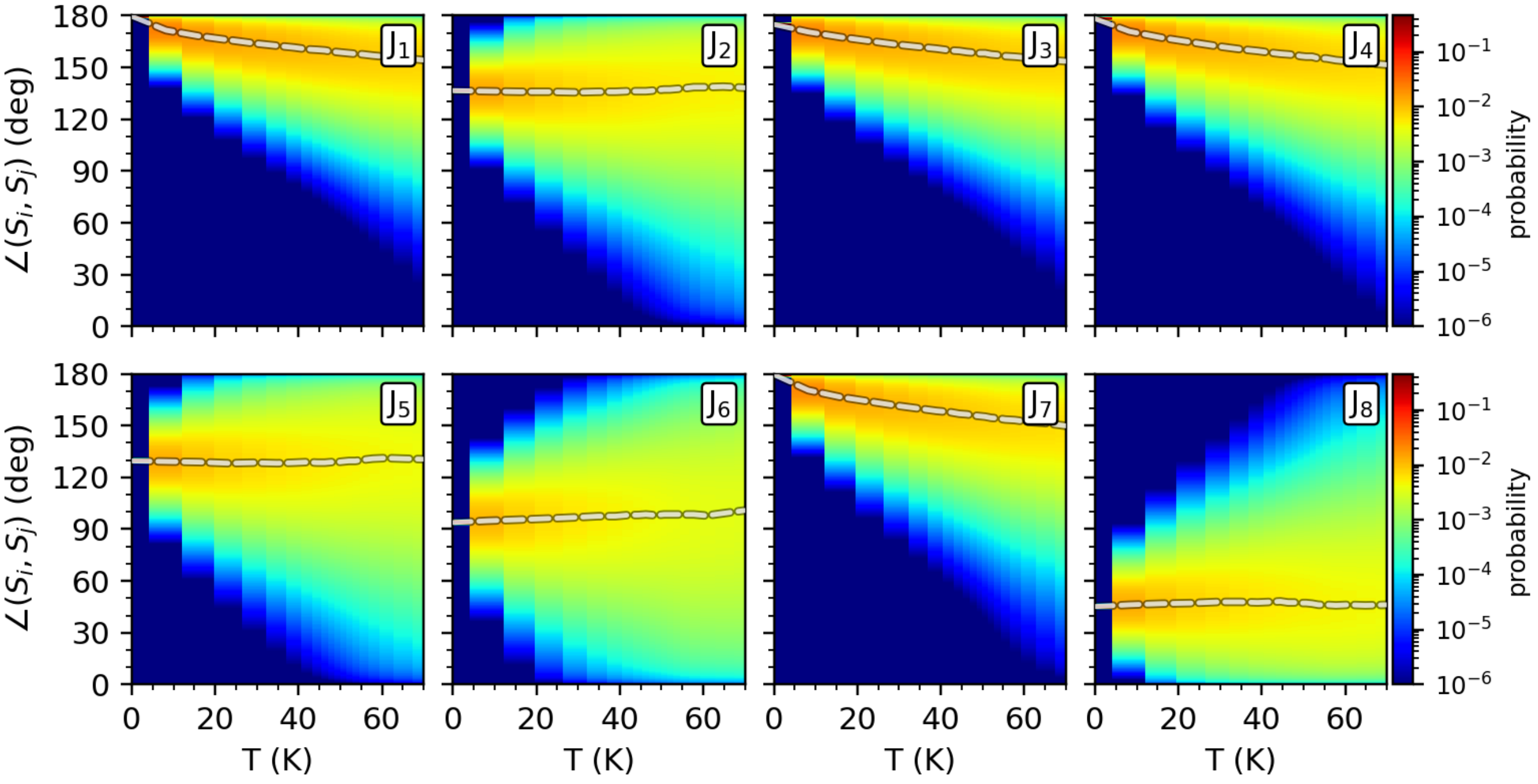


Supplementary Figure 4. Bond angles of $CsBaFe_3F_{12}$ as a function of temperature simulated with cMC. The dashed line traces the angle with maximum probability for each exchange interaction $J_i$. The plots are based on a $L = 24$ supercell.

Supplementary Table 3. Number of measurements $N$ used in the cMC simulations for different linear system sizes $L$.

| $CsBaFe_3F_{12}$ $L$ | $N$ |
|---|---|
| 8 | 600 000 |
| 12 | 200 000 |
| 16 | 60 000 |
| 20 | 33 000 |
| 24 | 20 000 |
| 32 | 10 000 |
| 36 | 15 000 |
| 40 | 15 000 |

| $CsBaCr_3F_{12}$ $L$ | $N$ |
|---|---|
| 12 | 100 000 |
| 16 | 55 000 |
| 20 | 22 000 |
| 24 | 18 000 |
| 28 | 13 000 |
| 32 | 9 500 |
| 36 | 25 000 |
| 40 | 25 000 |
| 42 | 48 000 |
| 44 | 44 000 |
| 48 | 25 000 |
| 50 | 28 800 |
| 52 | 28 000 |
| 56 | 34 000 |
| 64 | 27 000 |

$J_6$ are the triangular bonds. The data in Fig. 5 again show local-Neél configurations on the hexagons starting above the ordering temperature, while the triangular bonds order at much lower temperature. The evolution of the most probable angle on the latter is non-monotonous and qualitatively different from $CsBaFe_3F_{12}$, indicating that the two compounds mainly differ in how the triangular bonds interact with the local-Neél configuration on the hexagons.

## Supplementary Note 3. ANALYSIS OF ORDERING WAVEVECTORS IN THE EQUAL-TIME SPIN STRUCTURE FACTOR

Let $\mathbf{S}_{R\alpha}(T)$ be the classical spin on unit cell $R$ and basis site $\alpha$ at temperature $T$. The number of magnetic unit cells is $N_{\rm cell}$, the number of basis sites is $N_{\rm sub}$, and $N_{\rm site} = N_{\rm cell}N_{\rm sub}$. For a wave vector $\mathbf{q}$ in the simulation reciprocal grid, we first compute the sublattice-resolved spin Fourier amplitude

$$A_{\alpha\mu}(\mathbf{q},T) = \sum_R \exp[2\pi i\,\mathbf{q}\cdot\mathbf{r}_{R\alpha}]\,S_{R\alpha,\mu}(T), \tag{S1}$$

where $\mu = x, y, z$ labels the spin component. The coordinates $\mathbf{r}_{R\alpha}$ are the fractional coordinates in terms of the simulation cell, so that $\mathbf{q}\cdot\mathbf{r}_{R\alpha}$ is dimensionless.

For each $\mathbf{q}$, we form the $N_{\rm sub}\times N_{\rm sub}$ Hermitian matrix

$$C_{\alpha\beta}(\mathbf{q},T) = \frac{1}{N_{\rm site}}\sum_\mu A_{\alpha\mu}(\mathbf{q},T)A^*_{\beta\mu}(\mathbf{q},T). \tag{S2}$$

Its largest eigenvalue is denoted by $\lambda_{\max}(\mathbf{q},T)$. For a reduced ordering channel $\mathbf{q}_\nu$, we consider the set of symmetry-equivalent momenta $\mathcal{S}_\nu$, i.e. the set of reciprocal-grid wave vectors that fold onto $\mathbf{q}_\nu$. The quantity used to choose the reference temperature is

$$M(\mathbf{q}_\nu,T) = \left[\frac{2}{N_{\rm cell}}\max_{\mathbf{q}\in\mathcal{S}_\nu}\lambda_{\max}(\mathbf{q},T)\right]^{1/2}. \tag{S3}$$

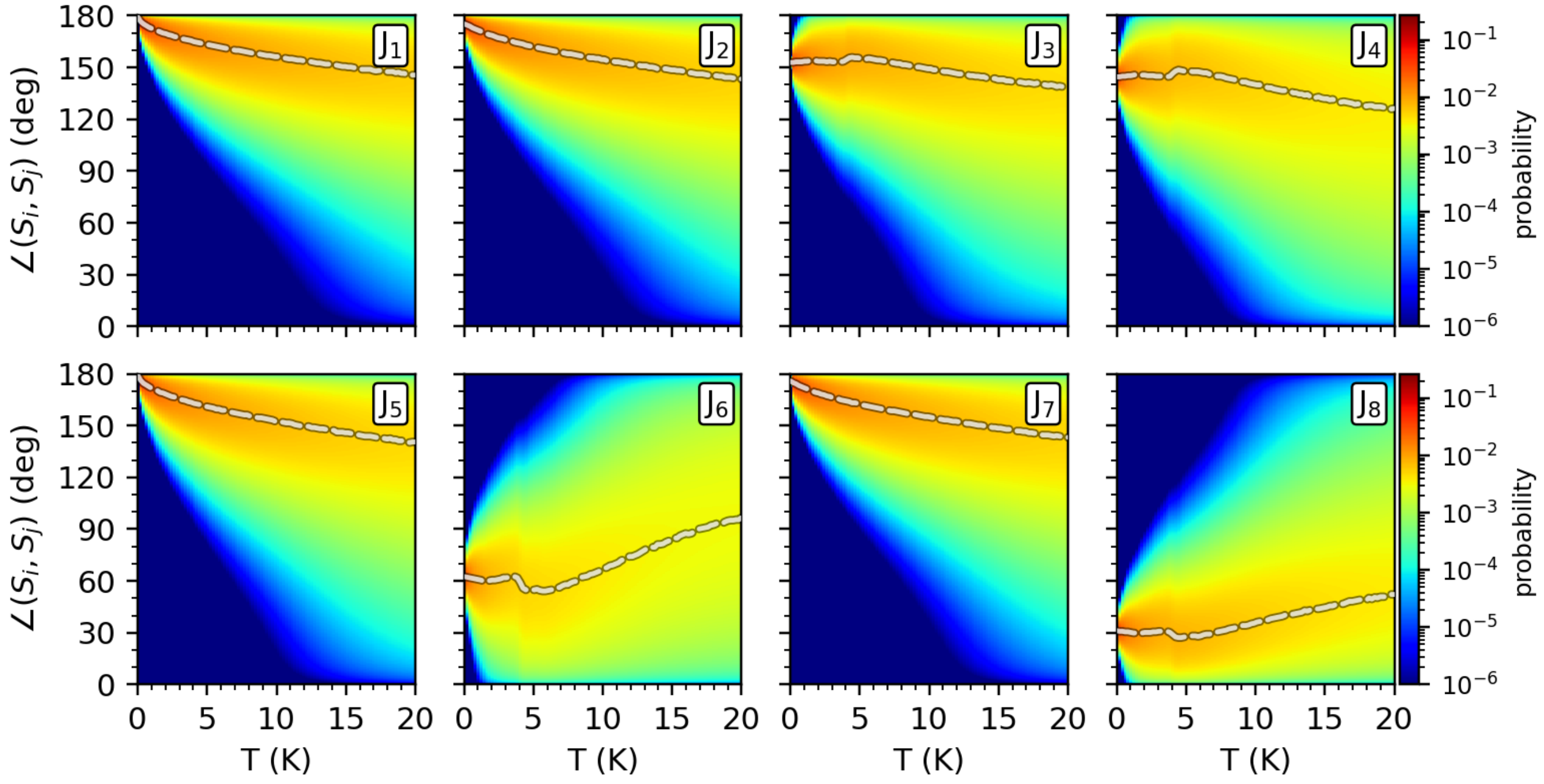


Supplementary Figure 5. Bond angles of $CsBaCr_3F_{12}$ as a function of temperature simulated with cMC. The dashed line traces the angle with maximum probability for each exchange interaction $J_i$. The plots are based on a $L = 40$ supercell.

The factor 2 is needed for proper normalization, since each spin configuration has the same Fourier weight at both $\mathbf{q}$ and $-\mathbf{q}$.

The reference channels, along with the temperatures, i.e. the temperature at which $M(\mathbf{q}_\nu, T)$ is maximal, and reference-state amplitudes are summarized in Table 2 of the main text.

For $CsBaCr_3F_{12}$ this analysis produces the four most relevant ordering vectors

$$\mathbf{q}_1 = 0.40(\mathbf{b}_1 + \mathbf{b}_2) - \mathbf{b}_1, \quad \mathbf{q}_2 = 0.425(\mathbf{b}_1 + \mathbf{b}_2) - \mathbf{b}_1,$$
$$\mathbf{q}_3 = 0.45(\mathbf{b}_1 + \mathbf{b}_2) - \mathbf{b}_1, \quad \mathbf{q}_4 = 0.475(\mathbf{b}_1 + \mathbf{b}_2) - \mathbf{b}_1, \tag{S4}$$

in reduced reciprocal-lattice coordinates. The corresponding reference temperatures are $0.010$ K, $1.736$ K, $4.557$ K, and $5.685$ K, respectively. For $CsBaFe_3F_{12}$ the same construction yields the single channel $\mathbf{q}_{\mathrm{Fe}} = 0.375(\mathbf{b}_1 + \mathbf{b}_2) - \mathbf{b}_1$.

The non-trivial sublattice part of each reference state is the leading eigenvector $u_\alpha(\mathbf{q})$ of Eq. (S2), evaluated at the corresponding reference temperature. Since the overall complex phase of an eigenvector is arbitrary, Table 4 fixes the gauge by making the $\alpha = 0$ component real and positive, so that $\theta_0 = 0$. For an ideal constant-moment single-$q$ spiral on symmetry-equivalent magnetic sites, one expects

$$u_\alpha = \frac{1}{\sqrt{N_{\mathrm{sub}}}} \exp(i\varphi_\alpha), \tag{S5}$$

so that all $|u_\alpha|$ are equal. The values in Table 4 are close to $1/\sqrt{12} \approx 0.289$, indicating that the extracted reference modes are well approximated by such a spiral, while retaining the small deviations present in the Monte Carlo snapshots.

For each channel $\nu$ and set of symmetry-equivalent momenta $\mathbf{q} \in \mathcal{S}_\nu$, we extract a reference mode from the spin configuration at $T_\nu^{\mathrm{ref}}$. Let $u_\alpha(\mathbf{q})$ be the normalized eigenvector of $C_{\alpha\beta}(\mathbf{q}, T_\nu^{\mathrm{ref}})$ corresponding to its largest eigenvalue. The associated spin-space polarization is

$$p_\mu(\mathbf{q}) = \sum_\alpha u_\alpha^*(\mathbf{q}) A_{\alpha\mu}(\mathbf{q}, T_\nu^{\mathrm{ref}}). \tag{S6}$$

The complex reference mode on the basis is then

$$B_{\alpha\mu}(\mathbf{q}) = u_\alpha(\mathbf{q})\, p_\mu(\mathbf{q}). \tag{S7}$$

At any temperature $T$, this mode defines a real-space spiral template

$$\mathcal{B}_{R\alpha,\mu}(\mathbf{q}) = B_{\alpha\mu}(\mathbf{q}) \exp[-2\pi i\, \mathbf{q} \cdot \mathbf{r}_{R\alpha}]\,. \tag{S8}$$

We write its real and imaginary parts as

$$X_{R\alpha,\mu}(\mathbf{q}) = \mathrm{Re}\, \mathcal{B}_{R\alpha,\mu}(\mathbf{q}),$$
$$Y_{R\alpha,\mu}(\mathbf{q}) = \mathrm{Im}\, \mathcal{B}_{R\alpha,\mu}(\mathbf{q}). \tag{S9}$$

A global spiral phase $\phi$ gives the real trial state

$$\mathcal{R}_{R\alpha,\mu}(\mathbf{q}, \phi) = \cos\phi\, X_{R\alpha,\mu}(\mathbf{q}) - \sin\phi\, Y_{R\alpha,\mu}(\mathbf{q}). \tag{S10}$$

The order parameter is the normalized overlap between the Monte Carlo spin configuration and the reference spiral,

Supplementary Table 4. Normalized leading eigenvectors used in the representative reference states. The numbered columns give the sublattice index $\alpha = 0, \ldots, 11$. For each reference state, the first row gives $|u_\alpha|$ and the second row gives the normalized phase $\theta_\alpha/\pi$, where $u_\alpha = |u_\alpha| \exp(i\theta_\alpha)$. Values are rounded to three decimals.

| case | quantity | sublattice index $\alpha$ | | | | | | | | | | | |
|---|---|---|---|---|---|---|---|---|---|---|---|---|---|
| | | 0 | 1 | 2 | 3 | 4 | 5 | 6 | 7 | 8 | 9 | 10 | 11 |
| Fe $\mathbf{q}_{\mathrm{Fe}} = 0.375(\mathbf{b}_1 + \mathbf{b}_2) - \mathbf{b}_1$ | $\lvert u_\alpha \rvert$ | 0.288 | 0.289 | 0.289 | 0.289 | 0.289 | 0.289 | 0.289 | 0.289 | 0.288 | 0.289 | 0.289 | 0.289 |
| | $\theta_\alpha/\pi$ | 0.000 | -0.612 | -0.734 | -0.122 | 0.929 | 0.826 | 0.337 | 0.440 | -0.284 | 0.339 | -0.450 | 0.928 |
| Cr $\mathbf{q}_1 = 0.400(\mathbf{b}_1 + \mathbf{b}_2) - \mathbf{b}_1$ | $\lvert u_\alpha \rvert$ | 0.293 | 0.293 | 0.292 | 0.292 | 0.285 | 0.285 | 0.284 | 0.284 | 0.290 | 0.289 | 0.288 | 0.288 |
| | $\theta_\alpha/\pi$ | 0.000 | 1.000 | 0.446 | -0.554 | 0.521 | -0.480 | -0.074 | 0.926 | -0.014 | 0.986 | 0.461 | -0.540 |
| Cr $\mathbf{q}_2 = 0.425(\mathbf{b}_1 + \mathbf{b}_2) - \mathbf{b}_1$ | $\lvert u_\alpha \rvert$ | 0.284 | 0.284 | 0.288 | 0.288 | 0.289 | 0.289 | 0.294 | 0.293 | 0.286 | 0.286 | 0.291 | 0.291 |
| | $\theta_\alpha/\pi$ | 0.000 | -0.998 | 0.413 | -0.586 | 0.488 | -0.512 | -0.073 | 0.927 | -0.067 | 0.933 | 0.482 | -0.518 |
| Cr $\mathbf{q}_3 = 0.450(\mathbf{b}_1 + \mathbf{b}_2) - \mathbf{b}_1$ | $\lvert u_\alpha \rvert$ | 0.273 | 0.277 | 0.283 | 0.284 | 0.295 | 0.293 | 0.301 | 0.303 | 0.285 | 0.284 | 0.291 | 0.292 |
| | $\theta_\alpha/\pi$ | 0.000 | -0.999 | 0.095 | -0.905 | 0.290 | -0.711 | -0.201 | 0.799 | -0.394 | 0.606 | 0.482 | -0.521 |
| Cr $\mathbf{q}_4 = 0.475(\mathbf{b}_1 + \mathbf{b}_2) - \mathbf{b}_1$ | $\lvert u_\alpha \rvert$ | 0.276 | 0.273 | 0.276 | 0.281 | 0.297 | 0.301 | 0.302 | 0.307 | 0.285 | 0.289 | 0.285 | 0.290 |
| | $\theta_\alpha/\pi$ | 0.000 | 0.998 | 0.210 | -0.790 | 0.323 | -0.678 | -0.117 | 0.884 | -0.339 | 0.667 | 0.543 | -0.459 |

maximized over a global phase $\phi$, a global proper spin rotation $\hat{R} \in SO(3)$, and the symmetry-equivalent momenta. For a fixed $\mathbf{q}$ and $\phi$, define

$$
\begin{aligned}
\Omega(\mathbf{q},\phi,T) &= \max_{\hat{R}\in SO(3)} \frac{N(\mathbf{q},\phi,T)}{D_{\mathcal{R}}(\mathbf{q},\phi)\, D_S(T)}, \\
N(\mathbf{q},\phi,T) &= \sum_{R,\alpha,\mu,\rho} (\hat{R})_{\mu\rho} \mathcal{R}_{R\alpha,\rho}(\mathbf{q},\phi) S_{R\alpha,\mu}(T), \\
D_{\mathcal{R}}(\mathbf{q},\phi) &= \left[\sum_{R,\alpha,\mu} \mathcal{R}_{R\alpha,\mu}(\mathbf{q},\phi)^2\right]^{1/2}, \\
D_S(T) &= \left[\sum_{R,\alpha,\mu} S_{R\alpha,\mu}(T)^2\right]^{1/2}.
\end{aligned}
\tag{S11}
$$

The maximization over $SO(3)$ is evaluated using the Kabsch solution of the orthogonal Procrustes problem. [1, 2] Concretely, for each trial phase we form the $3 \times 3$ covariance matrix

$$
K_{\rho\mu}(\mathbf{q},\phi,T) = \sum_{R,\alpha} \mathcal{R}_{R\alpha,\rho}(\mathbf{q},\phi) S_{R\alpha,\mu}(T). \tag{S12}
$$

Let

$$
K = U\ \mathrm{diag}(\sigma_1,\sigma_2,\sigma_3)\, V^T \tag{S13}
$$

with $\sigma_1 \geq \sigma_2 \geq \sigma_3 \geq 0$ be its singular-value decomposition. The determinant correction is

$$
\eta = \det(VU^T), \qquad \hat{R}_{\mathrm{opt}} = V\ \mathrm{diag}(1,1,\eta)\, U^T, \tag{S14}
$$

which ensures $\det \hat{R}_{\mathrm{opt}} = +1$. Therefore, the numerator after maximizing over proper rotations is

$$
\max_{\hat{R}\in SO(3)} N(\mathbf{q},\phi,T) = \sigma_1 + \sigma_2 + \eta\,\sigma_3. \tag{S15}
$$

If $\eta = -1$, the last singular value is subtracted rather than added; this excludes the improper reflection that would be allowed by an unconstrained $O(3)$ fit.

The component order parameter plotted in Fig. 4(c) and Fig. 6(e) of the main text is

$$
m(\mathbf{q}_\nu, T) = \max_{\mathbf{q}\in\mathcal{S}_\nu} \max_{\phi} \Omega(\mathbf{q},\phi,T). \tag{S16}
$$

The phase maximization is performed numerically. The calculation first samples $\phi$ uniformly on $[0, 2\pi)$, then refines the best interval by a one-dimensional golden-section search. Because of the normalization in Eq. (S11), each component satisfies $0 \leq m(\mathbf{q}_\nu, T) \leq 1$.

## Supplementary Note 4. ANALYTICAL EFFECTIVE HEXAGON-SPIN APPROXIMATION FOR THE RUBY LATTICE

Our cMC calculations show that both $CsBaFe_3F_{12}$ and $CsBaCr_3F_{12}$ develop local Néel correlations on the hexagon plaquettes of the ruby lattice well above the ordering temperature. Here, we develop an analytical approximation for the ruby lattice, in which we assume that hexagons show a rigid Néel order, so that the ordering vector and inter-hexagon spin pattern are decided by the triangular bonds, which couple these Néel ordered hexagons. The resulting analytical theory explains the origin of the different ordering wave vectors and the much smaller LT stiffness of $CsBaCr_3F_{12}$. It is not meant to replace the full 12-site LT calculation quantitatively: the projection keeps only the effective hexagon-spin vectors and the triangular couplings, so absolute stiffnesses differ from the full model, while the trend and microscopic mechanism are transparent.

Let $\mathbf{R}$ denote the Bravais-lattice vector locating the center of a rigid hexagon, and let $a = 1, \ldots, 6$ label the six spins around that hexagon. We assign an alternating sign $\eta_a = \pm 1$ to the six sites, with neighboring sites on the same hexagon carrying opposite signs. For a perfectly rigid Néel hexagon, all six microscopic spins are determined by a single three-component unit vector $\mathbf{n}_\mathbf{R}$,

$$\mathbf{S}_{\mathbf{R}a} \simeq \eta_a \mathbf{n}_\mathbf{R}, \qquad |\mathbf{n}_\mathbf{R}| = 1, \tag{S17}$$

in the unit-spin convention. Equivalently, $\mathbf{n}_\mathbf{R}$ is the staggered magnetization axis of the hexagon centered at $\mathbf{R}$,

$$\mathbf{n}_\mathbf{R} \simeq \frac{1}{6} \sum_{a=1}^{6} \eta_a \mathbf{S}_{\mathbf{R}a}, \tag{S18}$$

followed by normalization to unit length in the hexagon-spin approximation. In this sense $\mathbf{n}_\mathbf{R}$ records the orientation of the local Néel pattern on one hexagon. The sign convention $\eta_a \to -\eta_a$, $\mathbf{n}_\mathbf{R} \to -\mathbf{n}_\mathbf{R}$ is a gauge choice and leaves all physical spins unchanged.

The microscopic Heisenberg Hamiltonian is

$$H = \sum_{i<j} J_{ij}\, \mathbf{S}_i \cdot \mathbf{S}_j. \tag{S19}$$

We split it into bonds belonging to the rigid hexagons and bonds connecting different hexagons. The former stabilize the local Néel texture and are not part of the projected model. The latter are the triangular bonds. For one triangular bond connecting site $a$ on the hexagon at $\mathbf{R}$ to site $b$ on the hexagon at $\mathbf{R}'$, the effective hexagon-spin substitution gives

$$J^{\text{tri}}_{\mathbf{R}a,\mathbf{R}'b}\, \mathbf{S}_{\mathbf{R}a} \cdot \mathbf{S}_{\mathbf{R}'b} \simeq J^{\text{tri}}_{\mathbf{R}a,\mathbf{R}'b}\, \eta_a \eta_b\, \mathbf{n}_\mathbf{R} \cdot \mathbf{n}_{\mathbf{R}'} . \tag{S20}$$

All microscopic triangular bonds connecting the same pair of hexagons are then summed. This gives the effective Hamiltonian

$$H_{\text{eff}} = \frac{1}{2} \sum_{\mathbf{R},\mathbf{R}'} K_{\mathbf{R},\mathbf{R}'}\, \mathbf{n}_\mathbf{R} \cdot \mathbf{n}_{\mathbf{R}'} \quad , \tag{S21}$$

with

$$K_{\mathbf{R},\mathbf{R}'} = \sum_{a,b} J^{\text{tri}}_{\mathbf{R}a,\mathbf{R}'b}\, \eta_a \eta_b \quad . \tag{S22}$$

Here the sum over $a, b$ includes only retained triangular microscopic bonds connecting the hexagons at $\mathbf{R}$ and $\mathbf{R}'$. The prefactor $1/2$ in Eq. (S21) is needed for the explicitly symmetrized sum over ordered pairs $\mathbf{R}, \mathbf{R}'$.

The factor $\eta_a \eta_b$ is essential: even if all microscopic triangular couplings are antiferromagnetic, their projection onto the staggered hexagon variables can enhance, reduce, or cancel the effective inter-hexagon coupling. For example, two antiferromagnetic microscopic bonds of equal strength can cancel in the projected model if they connect opposite combinations of the staggered signs.

In the projection we have left open how the Néel vectors of adjacent hexagons are actually oriented. However, the effective couplings $K_{\mathbf{R},\mathbf{R}'}$ depend on the canting angle via the dot product of local Néel vectors $\mathbf{n}_\mathbf{R} \cdot \mathbf{n}_{\mathbf{R}'}$. The true canting angle can either be found by minimizing the total energy or by extracting the spatial pattern associated with the ordering vector in a Luttinger-Tisza framework.

The projected Hamiltonian now only contains the triangular couplings of the standard ruby lattice. For $CsBaFe_3F_{12}$ and $CsBaCr_3F_{12}$ this means that we have also discarded all inter-layer couplings, so that the unit cell reduces from 12 sites in two different layers to 6 sites in a single decoupled layer.

The real-space projected Hamiltonian is then

$$H_{\text{eff}} = \frac{1}{2} \sum_{\mathbf{R},\mathbf{R}'} K(\mathbf{R}' - \mathbf{R})\, \mathbf{n}_\mathbf{R} \cdot \mathbf{n}_{\mathbf{R}'} \quad . \tag{S23}$$

$\mathbf{n}_\mathbf{R}$ is the local Néel vector of the hexagon centred at $\mathbf{R}$. The quantity $K(\Delta\mathbf{R})$ is the projected coupling between that hexagon and the hexagon displaced by $\Delta\mathbf{R} = \mathbf{R}' - \mathbf{R}$. It is obtained from Eq. (S22) by summing all retained triangular microscopic bonds connecting two rigid hexagons within the same layer. The factor $1/2$ again avoids double counting after symmetrizing the bond list.

We use the Fourier convention

$$\mathbf{n}_\mathbf{R} = \frac{1}{\sqrt{N}} \sum_{\mathbf{q}} e^{2\pi i \mathbf{q}\cdot\mathbf{R}} \mathbf{n}_\mathbf{q}, \quad \mathbf{n}_\mathbf{q} = \frac{1}{\sqrt{N}} \sum_{\mathbf{R}} e^{-2\pi i \mathbf{q}\cdot\mathbf{R}} \mathbf{n}_\mathbf{R} . \tag{S24}$$

$N$ is the number of single-layer unit cells and $\mathbf{q}$ is expressed in reduced reciprocal-lattice coordinates of that layer. Since $\mathbf{n}_\mathbf{R}$ is a three-component real spin-space vector, $\mathbf{n}_\mathbf{q}$ is a complex three-component Fourier amplitude and $\mathbf{n}_{-\mathbf{q}} = (\mathbf{n}_\mathbf{q})^*$.

Inserting Eq. (S24) into Eq. (S23) and using translational invariance gives

$$H_{\text{eff}} = \frac{1}{2} \sum_{\mathbf{q}} K(\mathbf{q})\, \mathbf{n}_{-\mathbf{q}} \cdot \mathbf{n}_\mathbf{q} \quad , \tag{S25}$$

with the Fourier-transformed projected interaction

$$K(\mathbf{q}) = \sum_{\Delta\mathbf{R}} K(\Delta\mathbf{R})\, e^{2\pi i \mathbf{q}\cdot\Delta\mathbf{R}} \quad . \tag{S26}$$

If a projected bond connects $\mathbf{R}$ to $\mathbf{R} + \Delta\mathbf{R}$ with coupling $K(\Delta\mathbf{R})$, the symmetrized Hamiltonian also contains the reverse bond from $\mathbf{R} + \Delta\mathbf{R}$ back to $\mathbf{R}$, equivalently a term with displacement $-\Delta\mathbf{R}$. The real-space exchange is symmetric, $K(-\Delta\mathbf{R}) = K(\Delta\mathbf{R})$. The two Fourier contributions are therefore complex conjugates,

$$\begin{aligned} &K(\Delta\mathbf{R}) e^{2\pi i \mathbf{q}\cdot\Delta\mathbf{R}} + K(-\Delta\mathbf{R}) e^{-2\pi i \mathbf{q}\cdot\Delta\mathbf{R}} \\ =&2K(\Delta\mathbf{R}) \cos(2\pi \mathbf{q}\cdot\Delta\mathbf{R}) , \end{aligned} \tag{S27}$$

which is real. Thus $K(\mathbf{q})$ is a real scalar interaction. Therefore, the eigenvalue of the Luttinger-Tisza method is simply $\lambda(\mathbf{q}) \equiv K(\mathbf{q})$.

Supplementary Table 5. Effective hexagon-spin projection results in the unit-spin convention. The coarse-grained hexagon energy $\lambda(h_*)$ and reduced-coordinate stiffness $\kappa_h$ scale by $S^2 = (5/2)^2$ for $CsBaFe_3F_{12}$ and $S^2 = (3/2)^2$ for $CsBaCr_3F_{12}$. The canting angle $\beta_*$ is the neighboring-hexagon vector angle on the $B$-type links. The angle $\alpha_* = 360^\circ - 2\beta_*$ is the canting angle along the $A$-type links. In the column $\kappa_k$ ($\mathring{A}^2$K) we have converted $\kappa_h$ to physical wave-vector units and divided by six, because there are six microscopic spins per rigid hexagon; the physical-spin value is obtained by multiplying this column by $S^2$.

| compound | $A$ (K) | $B$ (K) | $B/(2A)$ | $h_*$ | $\beta_*$ | $\alpha_*$ | $\lambda(h_*)$ (K) | $\kappa_h$ (K) | $\kappa_k$ ($\mathring{A}^2$K) |
|---|---|---|---|---|---|---|---|---|---|
| $CsBaFe_3F_{12}$ | 44.080 | 58.590 | 0.665 | 0.365696 | 131.65° | 96.70° | -166.036 | 2590.9 | 273.13 |
| $CsBaCr_3F_{12}$ | 15.454 | 26.615 | 0.861 | 0.415113 | 149.44° | 61.12° | -76.745 | 420.6 | 43.38 |

The remaining triangular bonds are $J_2$, $J_5$, $J_6$ for $CsBaFe_3F_{12}$ and $J_3$, $J_4$, $J_6$ for $CsBaCr_3F_{12}$. In both cases $J_6$ points along the diagonal of the resulting triangular lattice with displacement vector $\delta_A = \pm(1,1)$ in terms of the lattice vectors. The other two couplings are parallel to the lattice vectors with displacement $\delta_B = \pm(1,0), \pm(0,1)$ (see Fig. 2 in the main text). Inserting these into Eq. S26 yields

$$\lambda(\mathbf{q}) = \sum_{\delta_A} A\, e^{2\pi i \mathbf{q}\cdot\delta_A} + \sum_{\delta_B} B\, e^{2\pi i \mathbf{q}\cdot\delta_B} . \tag{S28}$$

Each pair gives a cosine, because the two phase factors are complex conjugates. Therefore, the full two-dimensional interaction is

$$\begin{aligned}\lambda(\mathbf{q}) = {}& 2A\cos[2\pi(q_1+q_2)] \\ & + 2B\cos(2\pi q_1) + 2B\cos(2\pi q_2) .\end{aligned} \tag{S29}$$

In $CsBaFe_3F_{12}$ and $CsBaCr_3F_{12}$ we have $A = 2J_6$. In $CsBaFe_3F_{12}$ we have $B = J_2 + J_5$ and in $CsBaCr_3F_{12}$ we have $B = J_3 + J_4$. From LT theory on the full Hamiltonian we already know that the minima in reciprocal space are located on the diagonal $\mathbf{q} = h(\mathbf{b}_1 + \mathbf{b}_2) - \mathbf{b}_1$, i.e. $q_1 = q_2 = h$. Therefore, we now restrict the momenta in Eq. S29 in this way:

$$\lambda(h) = 2A\cos(4\pi h) + 4B\cos(2\pi h) . \tag{S30}$$

In the interval $0 \le h \le 1/2$, the $B$-link phase advance lies in $[0,\pi]$, so we identify the $B$-bond canting angle as $\beta \equiv 2\pi h$ and rewrite Eq. (S30) as:

$$\lambda(\beta) = 2A\cos(2\beta) + 4B\cos(\beta) . \tag{S31}$$

Differentiating by the angle $\beta$, we obtain the first derivative,

$$\frac{\partial\lambda}{\partial\beta} = -4\sin\beta\,(2A\cos\beta + B) . \tag{S32}$$

Apart from the collinear endpoints $\sin\beta = 0$, the non-collinear stationary point satisfies

$$\cos\beta_* = -\frac{B}{2A},\ h_* = \frac{\beta_*}{2\pi},\ \beta_* = \arccos\left(-\frac{B}{2A}\right) . \tag{S33}$$

Equivalently, $\cos(2\pi h_*) = -B/(2A)$. This solution exists as an interior minimum for $B/(2A) < 1$ in the antiferromagnetic case considered here. The minimum value is

$$\lambda(h_*) = -2A - \frac{B^2}{A} . \tag{S34}$$

The curvature at this minimum can also be written analytically. Since

$$\frac{\partial^2\lambda(h)}{\partial h^2} = -32\pi^2 A\cos(4\pi h) - 16\pi^2 B\cos(2\pi h), \tag{S35}$$

using Eq. (S33) gives

$$\kappa_h = \frac{1}{3}\left.\frac{\partial^2\lambda(h)}{\partial h^2}\right|_{h=h_*} = \frac{32\pi^2 A}{3}\left[1 - \left(\frac{B}{2A}\right)^2\right] . \tag{S36}$$

In canting-angle units the corresponding stiffness is smaller by $(2\pi)^2$, i.e. $\kappa_h = (2\pi)^2\kappa_\beta$,

$$\kappa_\beta = \frac{1}{3}\left.\frac{\partial^2\lambda(\beta)}{\partial\beta^2}\right|_{\beta=\beta_*} = \frac{2}{3A}\left(4A^2 - B^2\right) . \tag{S37}$$

The factor $1/3$ is the usual classical LT normalization for three-component Heisenberg spins. This expression is useful because it exposes why $CsBaCr_3F_{12}$ has a softer energy landscape. The ratio $B/(2A)$ is closer to one in $CsBaCr_3F_{12}$, which moves the minimum closer to the Brillouin-zone boundary and, through Eq. (S36), reduces the curvature of $\lambda(h)$ near the minimum. In the limiting case $B/(2A) \to 1$, the quadratic stiffness vanishes.

This form also clarifies the fully symmetric triangular-coupling limit. If all three microscopic triangular couplings are equal to a common value $J_{\mathrm{tri}}$, then the projected sums become

$$A = 2J_{\mathrm{tri}},\ B = 2J_{\mathrm{tri}},\ \frac{B}{2A} = \frac{1}{2} . \tag{S38}$$

Using Eq. (S33) then gives $\cos\beta_* = -1/2$, i.e.

$$\beta_* = \frac{2\pi}{3} = 120^\circ . \tag{S39}$$

Thus the symmetric projected model recovers the usual $120^\circ$ compromise of an antiferromagnetic triangular network of local Néel vectors. The diagonal $A$-type links carry phase $2\beta_* = 240^\circ$, which folds to the $A$-bond canting angle $\alpha_* = 120^\circ$.

Our numerical results are presented in Table 5. The unit-spin ordering-vector stiffness is strongly reduced in $CsBaCr_3F_{12}$; including the physical spin lengths gives

$$\frac{\kappa_{k,\mathrm{Fe}}}{\kappa_{k,\mathrm{Cr}}}\frac{S^2_{\mathrm{Fe}}}{S^2_{\mathrm{Cr}}} = \frac{273.13\,\mathring{A}^2\mathrm{K}}{43.38\,\mathring{A}^2\mathrm{K}} \cdot \frac{(5/2)^2}{(3/2)^2} \approx 17.5 . \tag{S40}$$

This is the quantitative signature of the shallow energy landscape in $CsBaCr_3F_{12}$. The stiffness ratio is a bit larger than for the full LT calculation, because the effective hexagon-spin approximation overestimates the stiffness of $CsBaFe_3F_{12}$ by $\sim 60\%$.

The obtained bond angles can also be compared with the bond angle histograms in Figs. 4 and 5. For $CsBaFe_3F_{12}$ the simplified model finds an angle $\beta_* \sim 132°$ on the $J_2$ and $J_5$ bonds and an angle of $\alpha_* = 360° - 2\beta_* \approx 97°$ on the $J_6$ bond. The latter is a bit larger than in cMC, but the other values are very close to the most likely cMC bond angle (see Fig. 4). For $CsBaCr_3F_{12}$ we find an angle $\beta_* \sim 149°$ on the $J_3$ and $J_4$ bonds and an angle of $\alpha_* = 360° - 2\beta_* \approx 61°$ on the $J_6$ bond. Both values almost exactly reproduce our cMC findings (see Fig. 5).



\end{document}